\documentclass[12pt]{iopart}
\usepackage{iopams}
\usepackage{graphicx}

\begin{document}
\title[Phase transitions in multi-cut matrix models and Whitham hierarchies]{Phase transitions in multi-cut matrix models and matched solutions of Whitham hierarchies}
\author{ Gabriel \'Alvarez$^1$, Luis Mart\'{\i}nez Alonso$^1$, Elena Medina$^2$}
\address{$^1$ Departamento de F\'{\i}sica Te\'orica II,
         Facultad de Ciencias F\'{\i}sicas,
         Universidad Complutense,
         28040 Madrid, Spain}
\address{$^2$ Facultad de Ciencias,
         Universidad de C\'adiz,
         11510 Puerto Real, Spain}
\begin{abstract}
We present a method to study phase transitions in the large $N$ limit of matrix models using matched solutions of Whitham hierarchies. The endpoints of the eigenvalue spectrum as functions of the temperature are characterized both as solutions of hodograph equations and as solutions of a system of ordinary differential equations.  In particular we show that the free energy of the matrix model is the quasiclassical $\tau$-function of the associated hierarchy, and that critical processes in which the number of cuts changes in one unit are third-order phase transitions described by $C^1$ matched solutions of Whitham hierarchies. The method is illustrated with the Bleher-Eynard model for the merging of two cuts. We show that this model involves also a birth of a cut.
\end{abstract}
\maketitle
\section{Introduction}
We consider the partition function of the unitary ensemble of random Hermitian matrices~\cite{met,gin}
\begin{equation}\label{0.1}
  Z_N = \int_{-\infty}^{+\infty}
            \cdots
            \int_{-\infty}^{+\infty}
            \exp\left(-\frac{N}{T}\sum_{i=1}^N V(\lambda_i)\right)\prod_{i<j}(\lambda_i-\lambda_j)^2\
            \rmd \lambda_1\cdots \rmd \lambda_N,
\end{equation}
where $T>0$ plays the role of a temperature, and where the potential $V(z)$ is a real polynomial of even degree
\begin{equation}
  V(z)=\sum_{n=1}^{2p} t_n z^n,
  \label{pot}
\end{equation}
with $V(0)=0$ and positive leading coefficient $t_{2p}>0$. We are interested in the critical processes which arise in the large $N$ limit as $T$ varies. It is known~\cite{dei0,dei} that for a given temperature the support $J$ of the eigenvalue density $\rho(x)$ is the union of a finite number of real intervals
\begin{equation}
  J = \bigcup_{j=1}^s (\beta_{2j-1},\beta_{2j}),\quad \beta_1<\beta_2<\cdots<\beta_{2\,s}.
\end{equation}
The  density and its support are uniquely determined by the following conditions~\cite{dei}:
\begin{eqnarray}
  \label{2a}
  V(x) - 2 T \int_{-\infty}^\infty \log|x-y| \rho(y) \rmd y = L,\quad\mbox{for $x\in\bar{J}$},\\
  \label{2b}
  V(x) - 2 T \int_{-\infty}^\infty \log|x-y| \rho(y) \rmd y\geq L,\quad \mbox{for $x\notin\bar{J}$},
\end{eqnarray}
for some real constant $L$, together with the normalization
\begin{equation}
  \label{norm}
  \int_J \rho(x) \rmd x = 1.
\end{equation}
These conditions mean that the infimum of the functional
\begin{equation}
  E_V[\phi] = \int_{-\infty}^\infty V(x) \phi(x) \rmd x - T \int_{-\infty}^\infty\int_{-\infty}^\infty\log|x-y| \phi(x) \phi(y) \rmd x \rmd y,
\end{equation}
on the set of normalized functions $\phi(x)\geq 0$ in $(-\infty,\infty)$ with bounded support is attained at  $\phi=\rho$ (the constant $L$ is the Lagrange multiplier). In physical terms  $\rho(x)$ represents the equilibrium density of a normalized one-dimensional charge distribution in the presence of an external electrostatic potential $V(x)$.

The general picture of the model is that the $T>0$ semi-axis decomposes into a number of disjoint (not necessarily connected) subsets $I_s$ wherein the support of the eigenvalues has a fixed number $s$ of disjoint intervals.  The multicut case $s>1$ is essentially different from the one-cut case: in the latter, the partition function has a regular large $N$ topological expansion, whereas for $s>1$ no such regular expansion exists and oscillatory terms appear. Using a saddle-point argument it has been shown~\cite{bde}  that the asymptotic behavior of the partition function for $s>1$ is of the form
\begin{equation}
  \label{asbde}
  Z_N \sim \rme^{-N^2 F}\theta(N/2\pi\mathbf{\Omega}),
\end{equation}
where $\theta(\mathbf{z})$ is a Riemann theta function and
\begin{equation}
  F = - T\int_J V(x) \rho(x) \rmd x + T^2\int_J\int_J\log|x-y| \rho(x) \rho(y) \rmd x \rmd y.
  \label{foft}
\end{equation}

In this paper we  propose a method to study phase transitions $I_s\rightarrow I_{s\pm 1}$ such as the \emph{merging of two cuts}~\cite{b1,be,be2} or the \emph{birth of a cut}~\cite{ey2,bc}. Our method uses techniques of the theory of quasiclassical integrable systems~\cite{krich} (universal Whitham hierarchies) to  characterize  the endpoints $\beta_j$ of the support as functions of $T$ and of the coefficients $t_n$ of the potential $V$. Following Krichever's algebro-geometric scheme~\cite{krich,krich1,gra}, we derive hodograph type equations for the functions $\beta_j$ (which are determined from the expressions of the Abelian differentials of an underlying Riemann surface) and characterize the free energy~(\ref{foft}) as a quasiclassical $\tau$-function~\cite{krich}.  Although our approach is much inspired by well-known works on the analysis of the zero-dispersion limit of the Korteweg de Vries equation in terms of Whitham equations~\cite{dl1,dl2,dl3,dl4,dl4b,dl5,g1,g2,g3}, we provide an additional efficient tool to study solutions of Whitham equations in matrix models: we prove that the $\beta_j$ satisfy a system of first-order ordinary differential equations which is completely determined by the expressions of the Abelian differentials. In this approach the role of the complicated hodograph system of equations is reduced to supply initial data for the system of first-order differential equations. With this simplified technique we are also able to derive several important general properties of critical processes $I_s\rightarrow I_{s\pm 1}$ such as the $C^1$ matching of their corresponding solutions of the Whitham equations or the (third) order of the phase transitions that they represent \cite{be,be2,ey2,bc,gro}. 

To illustrate our method we consider the Bleher-Eynard potential~\cite{be,be2}
\begin{equation}
  \label{bem}
  V(x) = \frac{1}{4} x^4 - \frac{4}{3} c x^3 + (2 c^2-1) x^2 + 8 c x,
  \quad -1<c<1.
\end{equation}
This model was proposed as a concrete example of a third order phase transition $I_2\rightarrow I_{1}$ (merging of two cuts) at $T_c=1+4 c^2$. However our numerical and theoretical analyses of the corresponding system of first order ordinary differential equations shows that for $c \neq 0$ the model also exhibits another third order phase transition $I_1\rightarrow I_{2}$ (birth of a cut) at a certain temperature $\widetilde{T}_c$ with $0<\widetilde{T}_c<T_c$. We also  derive  the asymptotic behaviour of the functions $\beta_j(T)$ and  $\partial_T^n F(T) (n=0,\ldots,3)$ near both critical points $T_c$ and $\widetilde{T}_c$. For the merging of two cuts we obtain the Bleher-Eynard asymptotic expressions for $\beta_j(T)$ and calculate the magnitude of the jump of $\partial_T^3 F(T)$ at $T_c$. For the birth of a cut we find the same logarithmic behaviours as in the Eynard models~\cite{ey2}.

The layout of the paper is as follows. In section~2 we set up our notation and present the elements of the theory of Whitham hierarchies on hyperelliptic Riemann surfaces. Section~3 is devoted to the solutions of Whitham hierarchies underlying matrix models. In section~4 we discuss several general properties of $I_s\rightarrow I_{s\pm 1}$ phase transitions, whereas section~5 addresses the particular cases $I_1\leftrightarrow I_2$. We defer to Appendix~A a self-contained proof of the relation between the free energy of the matrix model and the $\tau$-function of the hierarchy, to Appendix~B some explicit expressions and asymptotic estimates of elliptic integrals, and to Appendix~C the derivation of asymptotic approximations to the solutions of the differential equations in neighborhoods of the critical points.
\section{Multi-cut matrix models and Whitham hierarchies}
We denote by $w(z)$ the double-valued function
\begin{equation}
  w(z)=\sqrt{\prod_{i=1}^{2s} (z-\beta_i)},
\end{equation}
by $w_1(z)$ the branch of $w(z)$ with asymptotic behavior $w_1(z)\sim z^s$ as $z\rightarrow\infty$, and by $w_{1,+}(x)$ the boundary value of $w_1(z)$ on the real line from above. It has been shown \cite{dei} that the eigenvalue density $\rho(x)$ takes the form
\begin{equation}\label{0.2}
  \rho(x)=\frac{h(x)}{2\pi i\,T}\,\,w_{1,+}(x),\quad x\in J,
\end{equation}
where
\begin{equation}\label{0.3}
  h(z)=\left(\frac{V'(z)}{w_1(z)}\right)_{\oplus},
\end{equation}
and where $(f(z))_{\oplus}$ denotes the polynomial part of $f(z)$ at $z=\infty$. As a consequence of~(\ref{2a})
the endpoints $\beta_j$ of $J$ must satisfy the equations~\cite{dei}
\begin{eqnarray}
\label{e1} & \int_{\beta_{2j}}^{\beta_{2j+1}} h(x) w_{1,+}(x) \rmd x=0,\quad j=1,\ldots,s-1,\\
\label{e2} &\oint_{\gamma}z^j \frac{V'(z)}{w_1(z)} \rmd z=0,\quad j=0,\ldots,s-1,
\end{eqnarray}
where $\gamma$ is a large counter clockwise oriented loop which encircles $\bar{J}$.
Moreover, the normalization condition~(\ref{norm}) gives
\begin{equation}
\label{e3}
\oint_{\gamma}h(z)\,w_{1}(z)\,\rmd z=-4\pi\rmi T.
\end{equation}
Equations~(\ref{e1})--(\ref{e3}) give $2 s$ conditions to determine the $2 s$ unknowns $\beta_1,\ldots,\beta_{2s}$.
However, it may happen that  for fixed $T$ and $V(x)$ there exist different values of $s$ for which the system~(\ref{e1})--(\ref{e3}) has a solution $\beta_1<\cdots<\beta_{2\,s}$. If this is the case, the additional conditions $\rho(x)>0$ for all $x\in J$, and~(\ref{2b}) must be used to characterize the unique admissible solution of the problem. In terms of the functions $h$ and $w_1$ equation~(\ref{2b}) can be written as~\cite{dei}
\begin{eqnarray}
  \label{des1}
  &\int_x^{\beta_1}     h(x') w_{1}(x') \rmd x'\leq 0,\quad \mbox{for $x<\beta_1$} ,\\
  \label{des2}
  &\int_{\beta_{2j}}^x  h(x') w_{1}(x') \rmd x'\geq 0,\quad \mbox{for $\beta_{2j}<x<\beta_{2j+1},\quad j=1,\ldots s-1$}\\
  \label{des3}
  &\int_{\beta_{2s}}^x h(x') w_{1}(x') \rmd x'\geq 0,\quad \mbox{for $x>\beta_{2s}$}.
\end{eqnarray}
In addition there is an important \emph{a priori} upper bound for the number $s$ of cuts from a given potential $V$ at any temperature $T>0$. Equation~(\ref{0.3}) shows that $h(x)$ is a polynomial of degree
\begin{equation}
  \deg h=\deg V - s - 1,
\end{equation}
while from~(\ref{e1}) it is clear that $h(x)$ has at least one zero on each interval $\beta_{2j}<x<\beta_{2j+1}$. Therefore $\mbox{deg}\,h\geq s-1$ and we have the following bound:
\begin{equation}\label{nice}
s\leq \frac{\mbox{deg}\, V}{2}.
\end{equation}
Finally, the eigenvalue density $\rho(x)$ is called \emph{regular} if
$h(x)\neq 0$ for all $x\in\bar{J}$ and the inequalities~(\ref{des1})--(\ref{des3}) are strict. Otherwise $\rho(x)$ is called singular.

A basic property of multi-cut matrix models in the large $N$ limit is that~(\ref{e1})--(\ref{e3}), which determine the endpoints of the support of the eigenvalues, admit a natural interpretation in terms of Whitham hierarchies of quasiclassical integrable systems. In this section we describe the elements of the theory of Whitham hierarchies on hyperelliptic Riemann surfaces which are required for this interpretation.
\subsection{Hyperelliptic Riemann surfaces}
Let us denote by $\Gamma$ the hyperelliptic Riemann surface  associated with the curve
\begin{equation}
  \label{1.1}
  w^2=\prod_{i=1}^{2s} (z-\beta_i).
\end{equation}
The two branches $w_1(z)$ and $w_2(z)=-w_1(z)$ of the function $w(z)$ characterize $\Gamma$ as a double-sheeted covering of the extended complex plane:
\begin{equation}
  \Gamma=\Gamma_1\cup\Gamma_2,\quad \Gamma_i=\{Q=(w_i(z),z)\in\Gamma\}.
\end{equation}
The following coordinate systems constitute an atlas for $\Gamma$:
\begin{eqnarray}
  k_{\infty_i}=z^{-1},\quad \mbox{for $Q$ near $\infty_i,\,i=1,2$ }\\
  k_j=\sqrt{z-\beta_j}, \quad \mbox{for $Q$ near $(0,\beta_j)$},\\
  k=z,\quad \mbox{for $Q$ near any $Q_0\not\in \{\infty_1,\,\infty_2,\,(0,\beta_j)\}$}.
\end{eqnarray}
where $j=1,\ldots,2\,s$. The homology basis $\{a_i,b_i\}_{i=1}^{s-1}$ of cycles in $\Gamma$ is defined as shown in figure~1.
\begin{figure}
  \includegraphics[width=10cm]{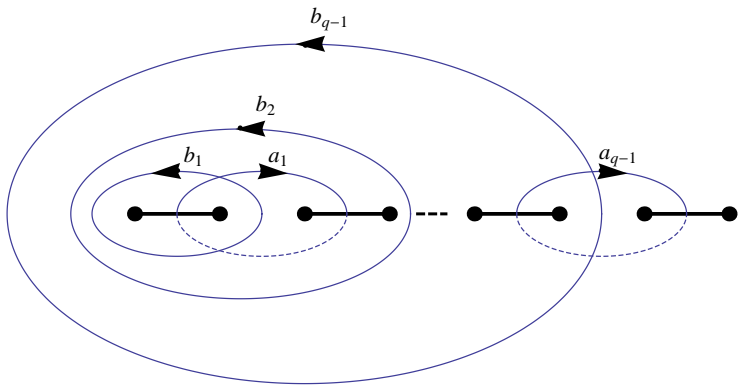}
  \caption{Homology basis.}
\end{figure}

The Whitham hierarchies describe certain deformations of the Riemann surface $\Gamma$ induced by functions $\beta_j=\beta_j(t_0,t_1,\ldots)$ in such a way that a suitable set $\{\rmd \Omega_k\}_{k=0}^\infty$ of Abelian differentials on $\Gamma$ can be derived from a ``potential'' $\rmd S$, i.e., $\rmd\Omega_k=\partial_{t_k}\rmd S$.

We denote by $\rmd\Omega_0$ the third kind normalized Abelian differential whose only poles are at
$\infty_1$ and $\infty_2$, and such that
\begin{equation}
  \label{o0}
\rmd \Omega_0(Q)=\left\{\begin{array}{ll}
                                        \displaystyle\left(\frac{1}{z}+\mathcal{O}(z^{-2})\right)\rmd z,&  Q\rightarrow \infty_1, \\
                                       \displaystyle\left(-\frac{1}{z}+\mathcal{O}(z^{-2})\right)\rmd z,& Q\rightarrow \infty_2,
  \end{array}\right.
\end{equation}
where here and hereafter it is assumed tacitly that $z=z(Q)$.
For $k\geq 1$ we denote by $\rmd \Omega_k$ the second kind normalized Abelian differential whose only pole is at
$\infty_1$, and such that
\begin{equation}\label{ok}
\rmd \Omega_k(Q)=
(k\,z^{k-1}+\mathcal{O}(z^{-2}))\rmd z,\quad Q\rightarrow \infty_1.
\end{equation}
Note that the usual normalization conditions
\begin{equation}\label{nc}
 \oint_{a_i} \rmd \Omega_k=0,\quad i=1,\ldots,s-1,
\end{equation}
determine the differentials $\rmd \Omega_k$ uniquely~\cite{spr,far}. It is easy to see that
\begin{equation}
  \label{diff1}
  \rmd \Omega_k=\left(\frac{k}{2}\,z^{k-1}+\frac{P_k(z)}{w(z)}\right)\rmd z,\quad k\geq 0,
\end{equation}
where $P_k(z)$ are polynomials of the form
\begin{equation}
  P_k(z)=\left(\delta_{k0}+\frac{k}{2}\right)(z^{k-1}\,w_1(z))_{\oplus}+\sum_{i=0}^{s-2}c_{k i}\,z^i
\end{equation}
and the coefficients $c_{k i}$ are determined by the normalization conditions~(\ref{nc}). We give explicitly the first polynomials for $s=1$ and $s=2$ which will be used in section~5.
\subsubsection*{Case $s=1$}
For $s=1$ we have
\begin{equation}
  \label{q1}
  P_k(z) = \left(\delta_{k0}+\frac{k}{2}\right)\left(z^{k-1}\,\sqrt{(z-\beta_1)(z-\beta_2)}\right)_{\oplus},
\end{equation}
so that  the first few polynomials are
\begin{eqnarray}
  \label{zero}
   &P_0(z)=1,\\
   &P_1(z)=\frac{1}{2}\,z-\frac{1}{4}\,(\beta_1+\beta_2),\\
   & P_2(z)=z^2-\frac{1}{2}\,(\beta_1+\beta_2)\,z-\frac{1}{8}\,(\beta_1-\beta_2)^2.
\end{eqnarray}
\subsubsection*{Case $s=2$}
For $s=2$ we have
\begin{equation}
  \label{q2}
  P_k(z) = \left(\delta_{k0}+\frac{k}{2}\right)
                 \left(z^{k-1}\sqrt{(z-\beta_1)(z-\beta_2)(z-\beta_3)(z-\beta_4)}\right)_{\oplus}+c_{k 0},
\end{equation}
where the coefficients $c_{k 0}$ are determined by the normalization condition
\begin{equation}
  \int_{\beta_2}^{\beta_3}\,\frac{P_k(x)}{\sqrt{(x-\beta_1)(x-\beta_2)(x-\beta_3)(x-\beta_4)}}\rmd x=0.
\end{equation}
Using the elliptic integrals in Appendix~B we find
\begin{eqnarray}
\label{el1} P_0(z)=&z-\beta_4+(\beta_4-\beta_3)\frac{\Pi(r,s)}{K(s)},\\
\label{elll} P_1(z)=&\frac{1}{2}z^2-\frac{1}{4}(\beta_1+\beta_2+\beta_3+\beta_4)\,z
                 + \frac{1}{4}\,(\beta_1\,\beta_4+\beta_2\,\beta_3)\nonumber\\
                 & {}+\frac{1}{4}\,(\beta_4-\beta_2)(\beta_3-\beta_1) \frac{E(s)}{K(s)},
\end{eqnarray}
where $K(s)$, $E(s)$ and $\Pi(r,s)$ are the complete elliptic integrals of first, second and third class, respectively, and
\begin{equation}
r=\frac{\beta_3-\beta_2}{\beta_4-\beta_2},\quad s=\frac{(\beta_4-\beta_1)(\beta_3-\beta_2)}{(\beta_3-\beta_1)(\beta_4-\beta_2)}.
\end{equation}

\begin{figure}
  \includegraphics[width=6cm]{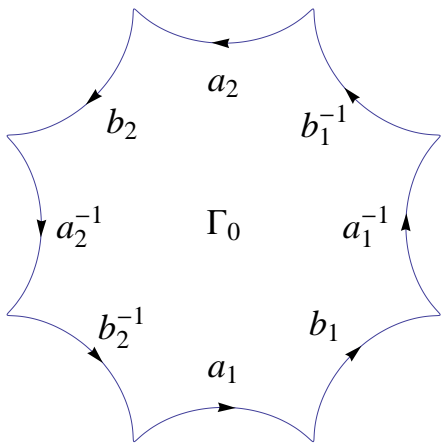}
  \caption{Fundamental domain.}
\end{figure}
Our next step is to define Abelian integrals $\Omega_k$ associated with the preceding Abelian differentials in such a way that $(\partial_z\,\Omega_k)\rmd z=\rmd \Omega_k$. We first deal with the generic case $\{\rmd\Omega_k\}_{k=1}^\infty$ and then consider separately the special case of $\rmd\Omega_0$. If we let all the basic cycles $\{a_i,b_i\}_{i=1}^{s-1}$ pass through some fixed point and cut  $\Gamma$ along these cycles we determine a simply-connected domain $\Gamma_0$, the so called fundamental domain of $\Gamma$ (figure~2). Thus,  by restricting our integration paths to $\Gamma_0-\{\infty_1\}$ we can define single-valued Abelian integrals
\begin{equation}\label{ak}
\Omega_k(Q)=\int_{\infty_2}^Q\,\rmd \Omega_k+L_k,\quad k\geq 1,
\end{equation}
where the $z$-independent term $L_k$ is
\begin{equation}\label{uk}
L_k=\lim_{Q\rightarrow \infty_1}( z^k-\int_{\infty_2}^Q\,\rmd \Omega_k).
\end{equation}
Note that from~(\ref{ok}) it follows that
\begin{equation}
  \int_{\infty_2}^Q \rmd \Omega_k=z^k+\mathcal{O}(1),\quad Q\rightarrow \infty_1,\quad k\geq 1
\end{equation}
and therefore
\begin{equation}
  \label{ake}
  \Omega_k(Q) = \left\{
                            \begin{array}{ll}
                             z^k+\mathcal{O}(z^{-1}),& Q\rightarrow \infty_1,\\
                             L_k+\mathcal{O}(z^{-1}),& Q\rightarrow \infty_2,
                             \end{array}
                           \right. k\geq 1.
\end{equation}

\begin{figure}
  \includegraphics[width=6cm]{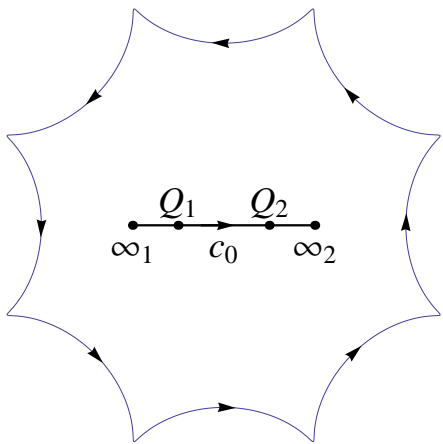}
  \caption{Path $c_0$.}
\end{figure}
In order to define a single-valued Abelian integral associated with $\rmd \Omega_0$ we fix a path $c_0$  in $\Gamma_0$ connecting $\infty_1$ with $\infty_2$, and a complex number $z_0$ such that both regular points $Q_1=(w_1(z_0),z_0)$ and $Q_2=(w_2(z_0),z_0)$ are in $c_0$ (figure~3). In this way we define the single-valued integral
\begin{equation}\label{a0}
  \Omega_0(Q) = \int_{\infty_2}^Q
                            \left(\rmd \Omega_0+\frac{\rmd z}{z-z_0}\right)+L_0-\log\,(z-z_0),
\end{equation}
where the integration path is in $\Gamma_0-c_0$, the constant term $L_0$ is defined by
\begin{equation}\label{u0}
L_0=\lim_{Q\rightarrow \infty_1}\left(2\,\log z-\int_{\infty_2}^Q\left(\rmd \Omega_0+\frac{\rmd z}{z-z_0}\right)\right),
\end{equation}
and $\log$ stands for the principal branch of the logarithm. Note that with this construction
\begin{equation}
  \rmd \Omega_0+\frac{\rmd z}{z-z_0}=\mathcal{O}(z^{-2})\,\rmd z,\quad Q\rightarrow \infty_2.
\end{equation}
Therefore the integral in~(\ref{a0}) is convergent and path-independent in $\Gamma_0-c_0$ because all the poles of the integrand are in $c_0$ and consequently the sum of the corresponding residues vanishes. Furthermore,
\begin{equation}
  \frac{\partial}{\partial z_0}\int_{\infty_2}^Q
   \left(\rmd \Omega_0+\frac{\rmd z}{z-z_0}\right)=-\frac{1}{z-z_0},\quad \frac{\partial}{\partial z_0} L_0=0,
\end{equation}
and consequently $\Omega_0(Q)$ is independent of $z_0$. It follows easily that
\begin{equation}
  \label{a0e}
  \Omega_0(Q) = \left\{\begin{array}{ll}
                                    \log\,z+\mathcal{O}(z^{-1}),& Q\rightarrow \infty_1,\\
                                   -\log\,z+L_0+\mathcal{O}(z^{-1}),& Q\rightarrow \infty_2.
                                   \end{array}\right.
\end{equation}
\subsection{Whitham hierarchies}
As we anticipated in the preceding section we consider deformations of the Riemann surface $\Gamma$ induced by functions
$\beta_j=\beta_j(t_0,t_1,\ldots)$ such that
\begin{equation}
  \label{w1}
  \partial_{t_j}\rmd \Omega_k=\partial_{t_k}\rmd \Omega_j,\quad j,k\geq 0.
\end{equation}
These equations are the Whitham equations for the differentials $\rmd \Omega_k$ and are the integrability conditions for the existence of a meromorphic differential $\rmd S$ satisfying
\begin{equation}
  \label{es}
  \partial_{t_k} \rmd S=\rmd \Omega_k,\quad k\geq 0.
\end{equation}
Hereafter we will use the vector notation ${\bbeta}=(\beta_1,\ldots,\beta_{2s})$, and $\mathbf{t}=(t_0,t_1,t_2,\ldots)$, so that ${\bbeta}={\bbeta}(\mathbf{t})$.

Integrating~(\ref{es}) with respect to all the deformation parameters and enforcing the normalization conditions
\begin{equation}
  \label{1.3}
  \oint_{a_i}\,\rmd S=0,\quad i=1,\ldots,s-1,
\end{equation}
we infer the existence of a unique normalized differential $\rmd S$ whose only poles are at $\infty_1$ and $\infty_2$ with expansions of the form
\begin{equation}
  \label{dse}
  \fl
  \rmd S(\mathbf{t},Q) = \left\{\begin{array}{ll}
                                             \displaystyle
                                             \left(\sum_{k\geq 1}\,k\,t_k\,z^{k-1}+\frac{t_0}{z}+\mathcal{O}(z^{-2})\right)\rmd z,& Q\rightarrow \infty_1, \\
                                             \\
                                             \displaystyle
                                             \left(-\frac{t_0}{z}+\mathcal{O}(z^{-2})\right)\rmd z,& Q\rightarrow \infty_2.
                                             \end{array}\right.
\end{equation}

Our next goal is to pass from this differential to its corresponding Abelian integral. Thus, from~(\ref{ak}), (\ref{uk}), (\ref{a0}), (\ref{u0}) and~(\ref{w1}) we obtain
\begin{equation}
\partial_{t_j}\,L_k=\partial_{t_k}\, L_j,\quad  j,k\geq 0,
\end{equation}
and therefore
\begin{equation}\label{w3}
\partial_{t_j}\,\Omega_k=\partial_{t_k}\, \Omega_j,\quad j,k\geq 0,
\end{equation}
which are the Whitham equations for the Abelian integrals $\Omega_k(\mathbf{t},z)$. Consequently, if we define by 
\begin{equation}\label{sf}
S(\mathbf{t},Q)=\int_{\infty_2}^Q \left(\rmd S+t_0\frac{\rmd z}{z-z_0}\right)+L(\mathbf{t})-t_0\,\log(z-z_0)
\end{equation}
the Abelian integral on $\Gamma_0-c_0$   corresponding to the differential $\rmd S$ that satisfies the conditions~(\ref{1.3})--(\ref{dse}), and if $L$ is such that
\begin{equation}
\partial_{t_k}\, L=L_k,\quad k\geq 0,
\end{equation}
then~(\ref{es}) implies at once that
\begin{equation}\label{w4}
\partial_{t_k}\, S(\mathbf{t},z)=\Omega_k(\mathbf{t},z),\quad k\geq 0.
\end{equation}
Moreover, by adjusting a constant term in $L$ we deduce that the function $S$ has the asymptotic expansions
\begin{equation}
  \label{sea}
  S(\mathbf{t},Q) = \left\{\begin{array}{ll}
                              \displaystyle
                              \sum_{k\geq 1}\,t_k\,z^{k}+t_0\,\log z-\sum_{i\geq 1}\frac{v_{i+1}}{i\,z^i},& Q\rightarrow \infty_1, \\
                              \\
                              \displaystyle
                               v_1-t_0\,\log z+\sum_{i\geq 1}\frac{v_{i+1}}{i\,z^i},& Q\rightarrow \infty_2.
                              \end{array}\right.
\end{equation}

The most basic object of the Whitham hierarchy is the quasiclassical $\tau$-function $\tau(\mathbf{t})=\exp F(\mathbf{t})$. It
is characterized \cite{krich} by the equations
\begin{equation}\label{idtau}
v_{i+1}=\frac{\partial F}{\partial t_i},\quad i\geq 0,
\end{equation}
where the functions $v_i$ are the coefficients of the expansion of $S$ given in~(\ref{sea}).
We prove in Appendix~A that
\begin{equation}\label{tau}
\quad F(\mathbf{t})  = \frac{1}{4\pi\rmi}\oint_{\gamma}V(z) \rmd S(z) + \frac{1}{2}t_0\,v_{1}
                                = \frac{1}{2}\sum_{i\geq 0}t_i\,v_{i+1}(\mathbf{t}),
\end{equation}
where $\gamma$ is a large counter clockwise oriented loop which encircles $\bar{J}$ in $\Gamma_1$.
Note that the calculation of the $\tau$-function requires knowledge of the $v_1$ coefficient, which in turn requires the construction of Abelian integrals and not just of Abelian differentials. In the next section we will see that $F$ is precisely the free energy in the leading nonoscillatory factor in~(\ref{asbde}), and that $v_1$ is the Lagrange multiplier in~(\ref{2a}).
\section{Solutions of Whitham hierarchies underlying multi-cut matrix models}
The purpose of this section is to prove that given a family of matrix models with a fixed number $s$ of cuts, then the differential
\begin{equation}
  \label{1.2}
  \rmd S = \frac{1}{2} \left(V'(z)+\left(\frac{V'(z)}{w_1(z)}\right)_{\oplus} w(z)\right) \rmd z
             = \frac{1}{2} \left(V'(z)+h(z) w(z)\right) \rmd z
\end{equation}
is the normalized solution with asymptotic behaviour~(\ref{dse}) of Whitham's equations for a finite number of deformation parameters $t_0$, $t_1$, \ldots, $t_{2p}$.  Here $t_0=-T$ and $t_1,\ldots,t_{2p}$ are the coupling constants in the potential~(\ref{pot}).

First note that~(\ref{e1}) simply means
\begin{equation}
  \oint_{a_i} \rmd S = 0,\quad i=1,\ldots,s-1.
\end{equation}
If we write $\rmd S$ in the form
\begin{equation}
  \rmd S = \frac{1}{2}
                \left( V'(z) + \frac{V'(z)}{w_1(z)}w(z) - \left(\frac{V'(z)}{w_1(z)}\right)_{\ominus} w(z) \right) \rmd z,
\end{equation}
where $(f(z))_\ominus = f(z) - (f(z))_\oplus$, and recall that $w_1(z)\sim z^s$ as $z\rightarrow \infty$, then it is obvious that~(\ref{e2}) and~(\ref{e3}) imply
\begin{equation}
 \left(\frac{V'(z)}{w_1(z)}\right)_{\ominus} w_1(z) = -\frac{2 t_0}{z}+\mathcal{O}(z^{-2}),\quad z\rightarrow \infty,
\end{equation}
or, equivalently, that $\rmd S$ has the asymptotic expansions
\begin{equation}
  \label{1.4}
  \rmd S(Q) = \left\{\begin{array}{ll}
                      \displaystyle\left(\sum_{k=1}^{2p}k\,t_k\,z^{k-1}+\frac{t_0}{z}+\mathcal{O}(z^{-2})\right)\rmd z,
                      \quad\mbox{as $Q\rightarrow \infty_1$},\\
                      \\
                      \displaystyle\left(-\frac{t_0}{z}+\mathcal{O}(z^{-2})\right)\rmd z,
                      \quad \mbox{as $Q\rightarrow \infty_2$}.
                     \end{array}\right.
\end{equation}
Moreover, in terms of the local coordinate $k_i$ near a point $(0,\beta_i)$ we have that
\begin{equation}
  z=k_i^2+\beta_i,\quad \rmd z= 2\,k_i \rmd k_i,
\end{equation}
so that
\begin{equation}
  \label{sb}
  \rmd S = \mathcal{O}(k_i) \rmd k_i,\quad k_i\rightarrow 0.
\end{equation}
Hence the differential $\rmd S$ has poles at $\infty_1$ and $\infty_2$ only, and from~(\ref{1.4}) it is clear that
\begin{equation}
  \label{ds1}
  \rmd S = \sum_{k=0}^{2p} t_k\,\rmd \Omega_k.
\end{equation}
Differentiating~(\ref{1.4}) with respect to $t_k$ $(k=1,\ldots,2p)$, it follows that
\begin{equation}
  \label{1.4bis}
  \partial_{t_k}\rmd S(Q) = \left\{\begin{array}{ll}
                                                \left(k z^{k-1} + \mathcal{O}(z^{-2})\right)\rmd z,
                                                \quad\mbox{as $Q\rightarrow \infty_1$},\\
                                                \\
                                                \mathcal{O}(z^{-2})\rmd z,
                                                \quad \mbox{as $Q\rightarrow \infty_2$}.
                                          \end{array}\right.     
\end{equation}
Furthermore, (\ref{sb}) shows that near $(0,\beta_j)$
\begin{equation}
  \label{sb1}
  \partial_{t_k} \rmd S = \mathcal{O}(1) \rmd k_j, \quad k_j\rightarrow 0.
\end{equation}
Hence, since the  $a_l$-periods of $\partial_{t_k} \rmd S$ vanish, $\rmd S$ satisfies the Whitham equations
\begin{equation}\label{wno}
  \partial_{t_k} \rmd S = \rmd \Omega_k,\quad k = 1,\ldots, 2p.
\end{equation}
The case $k=0$ can be proved similarly.

As a consequence of~(\ref{1.2}) and~(\ref{ds1}) we obtain an important identity which for later reference we state making explicit the dependence of $P$, $h$ and $w$ on the endpoints ${\bbeta}$:
\begin{equation}
  \label{id11}
  \sum_{k\geq 0} t_k P_k(z,{\bbeta}) = \frac{1}{2} h(z,{\bbeta}) w(z,{\bbeta})^2.
\end{equation}
\subsection{Hodograph equations}
If we set $z=\beta_i$ in~(\ref{id11}), we find that the solutions of the Whitham equations~(\ref{es}) associated with multi-cut matrix models satisfy the hodograph type equations
\begin{equation}
  \label{hod}
  \sum_{k=0}^{2p} t_k P_k(\beta_i,{\bbeta}) = 0,\quad i=1,\ldots,2\,s.
\end{equation}
These equations can be stated in a geometric formulation~\cite{krich1,gra} if we recall that in terms of the local coordinate $k_i(\mathbf{t})=\sqrt{z-\beta_i(\mathbf{t})}$ near $(0,\beta_i)$, we have that for all $k\geq 0$
\begin{equation}
  \label{ppn}
  \rmd \Omega_k = \left(
                               \frac{2 P_k(\beta_i,{\bbeta})}{\sqrt{\prod_{l\neq i}(\beta_i-\beta_l)}}+\mathcal{O}(k_i) \right)\rmd k_i,
\quad k_i \rightarrow 0.
\end{equation}
Hence, equations~(\ref{ds1}) and~(\ref{hod}) imply that
\begin{equation}
  \label{ce}
  \rmd S|_{(0,\beta_i)}=0,\quad i=1,\ldots,2\,s.
\end{equation}
Thus the solutions of~(\ref{hod}) determine zeros of the differential $\rmd S$. Moreover, equating the principal parts of $\partial_{t_1}\rmd \Omega_k$ and of $\partial_{t_k}\rmd \Omega_1$ at the poles $(0,\beta_i)$ and taking into account that $\partial_{t_n}\rmd k_i=(\partial_{t_n} \beta_i/2 k_i^2)\rmd k_i $, we find that~(\ref{hod}) represent the hodograph transform~\cite{ts} for a compatible set of diagonal hydrodynamical systems given by
\begin{equation}
   \label{w2}
\partial_{t_k} \beta_i=\frac{P_k(\beta_i,{\bbeta})}{P_1(\beta_i,{\bbeta})}\partial_{t_1} \beta_i,\quad \quad k\geq 0.
\end{equation}
These are the Whitham equations in hydrodynamic form.
\subsubsection{Example}
The simplest non-trivial example that we may consider is the quartic even potential~\cite{b1}:
\begin{equation}
  V(z) = \frac{1}{4} z^4 - z^2.
  \label{eq:bepotc0}
\end{equation}
Its associated hodograph equations~(\ref{hod}) are
\begin{equation}\label{hoda}
  \frac{1}{4}\,P_4(\beta_i,{\bbeta})-P_2(\beta_i,{\bbeta})-T\,P_0(\beta_i,{\bbeta})=0,\quad i=1,\ldots,2s.
\end{equation} 
According to the bound~(\ref{nice}), as $T$ changes there are only two possibilities for the number of cuts: $s=1$ or $s=2$. For $s=1$ the hodograph equations~(\ref{hod}) read
\begin{equation}
\fl
-256\,T+35\,\beta_1^4-5\,\beta_2^4-20\,\,\beta_1^3\,\beta_2-6\,
\beta_1^2\,\beta_2^2-4\,\beta_1\,\beta_2^3-96\,\beta_1^2+32\,\beta_2^2+
64\,\beta_1\,\beta_2=0,\quad \quad
\label{eq:hods1a}
\end{equation}
\begin{equation}
\fl
-256\,T+35\,\beta_2^4-5\,\beta_1^4-20\,\,\beta_2^3\,\beta_1-6\,
\beta_2^2\,\beta_1^2-4\,\beta_2\,\beta_1^3-96\,\beta_2^2+32\,\beta_1^2+
64\,\beta_2\,\beta_1=0,\quad \quad
\label{eq:hods1b}
\end{equation}
and the function $h(z)$ is given by 
\begin{equation}
h(z)=z^2+\frac{z}{2}(\beta_1+\beta_2)+\frac{1}{8}\,(3\,\beta_1^2+3\,\beta_2^2+2\,\beta_1\,\beta_2)-2.
\end{equation}
Imposing the symmetry condition $\beta_2 = - \beta_1$, the hodograph equations reduce to the single equation
\begin{equation}
  2 T + \beta_1^2 - \frac{3}{8} \beta_1^4 = 0,
\end{equation}
which leads to the solution
\begin{equation}
\label{se1}
 \beta_1 = - \frac{2}{\sqrt{3}}\sqrt{1+\sqrt{1+3 T}},\quad \beta_2 = \frac{2}{\sqrt{3}}\sqrt{1+\sqrt{1+3 T}},
\end{equation}
which is real and satisfies $\beta_1<\beta_2$ for all $T>0$. However,
the corresponding function $h(z)$ takes the form
\begin{equation}
h(z)=z^2+\frac{\beta_1^2}{2}-2=z^2+\frac{2}{3}\,(1+\sqrt{1+3\,T})-2,
\end{equation}
and, obviously, is strictly positive on the real axis only for $T>1$. Furthermore, we have 
\begin{equation}
w_{1,+}(x) = \left\{\begin{array}{ll}
                      \displaystyle
                      -|x^2-\beta_1^2|^{1/2},
                      \quad\mbox{for $x\leq\beta_1$},\\
                      \\
                      \displaystyle
                      i\,|x^2-\beta_1^2|^{1/2},
                      \quad \mbox{for $\beta_1\leq x\leq \beta_2$}
                      \\
                      \\
                      \displaystyle
                      |x^2-\beta_1^2|^{1/2},
                      \quad \mbox{for $\beta_2\leq x$}.
                     \end{array}\right.
\end{equation}
Hence for $T>1$ the functions~(\ref{se1}) are the endpoints of the support of eigenvalues, since $\rho(x)>0$ for all $x\in (\beta_1,\beta_2)$ and the inequalities~(\ref{des1})--(\ref{des3}) are strictly satisfied. 

If for $s=2$ we impose the symmetry relations $\beta_4 = - \beta_1$, $\beta_3 = - \beta_2$, the terms $c_{k0}$ with even $k$ in~(\ref{q2}) vanish and the hodograph equations~(\ref{hoda}) reduce to the system  
\begin{equation}
  \beta_1 \left(2 T - \frac{3}{8} \beta_1^4  -  \beta_2^2 + \frac{1}{8}\beta_2^4 + \beta_1^2 + \frac{1}{4}\beta_1^2\beta_2^2\right) = 0,
  \label{eq:hods2a}
\end{equation}
\begin{equation}
  \beta_2 \left(2 T +  \frac{1}{8}\beta_1^4  + \beta_2^2 - \frac{3}{8} \beta_2^4 - \beta_1^2 + \frac{1}{4}\beta_1^2\beta_2^2\right) = 0,
  \label{eq:hods2b}
\end{equation}
which supplies the following  solution of~(\ref{hod}):
\begin{equation}\label{bs2}
  \beta_1 = - \sqrt{2(1+\sqrt{T})},
  \quad
  \beta_2 = - \sqrt{2(1-\sqrt{T})},
\end{equation}
\begin{equation}\label{bs3}
  \beta_3 = \sqrt{2(1-\sqrt{T})},
  \quad
  \beta_4 = \sqrt{2(1+\sqrt{T})}.
\end{equation}
Moreover, in this case $h(z)=z$ and 
\begin{equation}
w_{1,+}(x) = \left\{\begin{array}{ll}
                      \displaystyle
                      |(x^2-\beta_1^2)(x^2-\beta_2^2)|^{1/2},
                      \quad\mbox{for $x\leq \beta_1$
                      or $x\geq \beta_4$ },\\
                      \\
                      \displaystyle
                      -i\,|(x^2-\beta_1^2)(x^2-\beta_2^2)|^{1/2},
                      \quad \mbox{for $\beta_1\leq x\leq \beta_2$},
                      \\                      \\
                      \displaystyle
                      -|(x^2-\beta_1^2)(x^2-\beta_2^2)|^{1/2},
                      \quad\mbox{for $\beta_2\leq x\leq \beta_3$},\\
                      \\
                                            \displaystyle
                      i\,|(x^2-\beta_1^2)(x^2-\beta_2^2)|^{1/2},
                      \quad \mbox{for $\beta_3 \leq x \leq \beta_4$}.
                                          \end{array}\right.
\end{equation}
Hence for $0<T<1$ the functions~(\ref{bs2})--(\ref{bs3}) are the endpoints of the support of eigenvalues, since $\rho(x)>0$ for all $x\in (\beta_1,\beta_2)\cup(\beta_3,\beta_4)$ and the inequalities~(\ref{des1})--(\ref{des3}) are strictly satisfied. Therefore there is a merging of two cuts at $T=1$. Note finally that the functions $(\beta_1,\beta_2)$ of the $s=1$ sector have a $C^1$-smooth matching with the functions $(\beta_1,\beta_4)$ of the $s=2$ sector at $T=1$. Indeed, for $s=1$ we have 
\begin{equation}
  \beta_1 = - 2 - \frac{1}{4} (T-1) + \frac{1}{16} (T-1)^2 + \mathcal{O}((T-1)^3).
\end{equation}
while for $s=2$
\begin{equation}
  \beta_1 = - 2 - \frac{1}{4} (T-1) + \frac{5}{64} (T-1)^2 + \mathcal{O}((T-1)^3).
\end{equation}
Notice that the matching is $C^1$ but not $C^2$.
\begin{figure}
  \includegraphics[width=10cm]{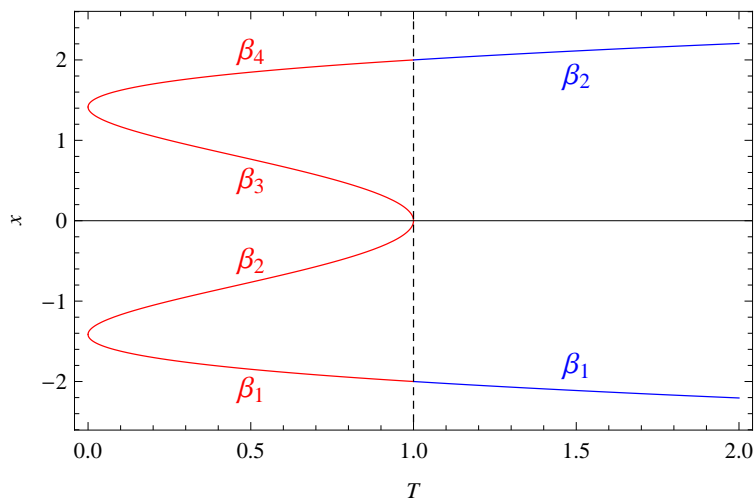}
  \caption{Evolution of the endpoints for the quartic even potential.}
\end{figure}
\subsection{An alternative approach to determine the eigenvalue support}
Except for the simplest examples, the system of hodograph equations~(\ref{hod}) is extremely difficult to solve, even by numerical methods. Our strategy to deal with this difficulty is to take advantage of a particular feature of the solutions  of the  Whitham hierarchies associated with matrix models: the functions $\beta_i(\mathbf{t})$ satisfy a system of first-order ordinary differential equations which will provide not only a practical numerical method but also analytical estimates for studying the asymptotic behaviour near critical temperatures.

From~(\ref{1.2}) and~(\ref{ds1}) it follows that
\begin{equation}
  \partial_{t_n} \log(h(z,{\bbeta}){w_1(z,{\bbeta})})
  =
  2\,\frac{P_n(z,{\bbeta})}{h(z,{\bbeta})\,w_1(z,{\bbeta})^2},
\end{equation}
where $n= 0,\ldots,2p$. Then, equating the residues of both sides at $z= \beta_i\,(i=1,\ldots,2s)$ we get
\begin{equation}\label{vel1a}
  \partial_{t_n}\beta_i
  = -\frac{4 P_n(\beta_i,{\bbeta})}{h(\beta_i,{\bbeta}) \prod_{j\neq i}(\beta_i-\beta_j)}.
\end{equation}
By construction this is a compatible system of first-order ordinary differential equations which is satisfied by the endpoints functions. Therefore, if we can find a solution ${\bbeta}(\mathbf{t}_0)$ of the hodograph equations~(\ref{hod}) at a certain value $\mathbf{t}_0$ 
of the  deformation parameters, then this ${\bbeta}(\mathbf{t}_0)$ will supply a set of initial data for solving the system~(\ref{vel1a}) and determine completely ${\bbeta}(\mathbf{t})$ in a region of constant $s$.

We stress that the first order ordinary differential equations~(\ref{vel1a}) in fact imply the full system of hydrodynamic Whitham partial differential equations~(\ref{w2}). Furthermore, if we denote by $x_1,\ldots,x_{s-1}$ the roots of $P_0(z,\bbeta)$, then the $n=0$ member of~(\ref{vel1a}) reduces to
\begin{equation}
      \partial_{T}\beta_i
  = \frac{4 \prod_{k=1}^{s-1}(\beta_i-x_k)}{h(\beta_i,{\bbeta}) \prod_{j\neq i}(\beta_i-\beta_j)} ,
\end{equation}
which are the Bleher-Eynard relations [A.13] in~\cite{be}.
\subsection{Relation between the $\tau$-function and the free energy}
We now establish the relation between the $\tau$-function and the nonoscillatory factor of the partition function of the matrix model in the large $N$ limit. From~(\ref{ds1}) and~(\ref{wno}) we deduce that
\begin{equation}
  \sum_{k=0}^{2p} t_k\,\partial_{t_i}\,\rmd \Omega_k=0.
\end{equation}
As a consequence the functions $L_k$ defined in~(\ref{uk}) and~(\ref{u0}) satisfy
\begin{equation}
  \partial_{t_n} L(\mathbf{t}) = L_n,\quad n=0,\ldots,2p,
\end{equation}
where
\begin{equation}
  \label{mult0}
  L(\mathbf{t}) = \sum_{k=0}^{2p} t_k\,L_k(\mathbf{t}).
\end{equation}
Thus the Abelian integral~(\ref{sf}) corresponding to the differential~(\ref{1.2}) is
\begin{equation}
  \label{sgood}
  S(\mathbf{t},Q) = \sum_{k=0}^{2p} t_k \Omega_k(\mathbf{t},Q),
\end{equation}
and has the asymptotic behaviour given by~(\ref{sea}) with
\begin{equation}
  \label{coefv1}
  v_1(\mathbf{t})=L(\mathbf{t}).
\end{equation}
In particular we get that $v_1$ satisfies
\begin{equation}
  \label{uve1a}
  \partial_{t_n}\,v_1=L_n,\quad n=0,\ldots,2p.
\end{equation}

To relate the function $S$ to the eigenvalue density $\rho(x)$ of the matrix model we prove that the restrictions $S_i$ of the function $S$ to the sheets $\Gamma_i$ $(i=1,2)$ of the Riemann surface can be written in the form
\begin{equation}
  \label{rest1}
  S_1(\mathbf{t},z) = V(\mathbf{t},z) + t_0 \int_{J} \rho(x) \log(z-x) \rmd x,
\end{equation}
\begin{equation}
  \label{rest2}
  S_2(\mathbf{t},z) = v_1(\mathbf{t}) - t_0 \int_{J} \rho(x) \log(z-x) \rmd x.
\end{equation}
We begin by noticing that the derivatives of $S_i$ with respect to $z$ are given by
\begin{equation}
  S'_1(z) = \frac{1}{2} \left( V'(z) + \left(\frac{V'(z)}{w_1(z)}\right)_{\oplus} w_1(z)\right),
\end{equation}
\begin{equation}
  S'_2(z) = V'(z) - S'_1(z).
\end{equation}
Thus, the boundary values of $S'_1(z)$ on the real axis satisfy
\begin{equation}
  S'_1(x+\rmi 0) - S'_1(x-\rmi 0) = \left\{\begin{array}{l}
                                          \displaystyle\left(\frac{V'(x)}{w_1(x)}\right)_{\oplus} w_1(x),\quad \mbox{for $x\in J$}\\
                                          \\
                                         0,\quad \mbox{for $x\in(-\infty,\infty)-J$}.
                                         \end{array}\right.
\end{equation}
Hence
\begin{equation}
  S'_1(z) = V'(z) + t_0 \int_{J}\frac{\rho(x)}{z-x} \rmd x,
\end{equation}
\begin{equation}
  S'_2(z) = -t_0 \int_{J}\frac{\rho(x)}{z-x} \rmd x.
\end{equation}
Therefore, taking into account the normalization of $\rho(x)$ and~(\ref{sea}), our equations~(\ref{rest1}) and~(\ref{rest2}) follow. Moreover since the function $S$ is continuous in $\Gamma_0-c_0$ (cf.~figure~3) we have
\begin{equation}
  S_1(x+\rmi 0) = S_2(x-\rmi 0),\quad \mbox{for all $x\in J$}.
\end{equation}
Thus, using~(\ref{rest1}), (\ref{rest2}) and
\begin{equation}
  \log(x-y+\rmi 0) + \log(x-y-\rmi 0) = 2 \log|x-y|,\quad (x\neq y),
\end{equation}
we obtain
\begin{equation}
  \label{multnn}
 v_1(\mathbf{t}) = V(x) + 2 t_0 \int_{J}\rho(y) \log|x-y| \rmd y,\quad x\in J.
\end{equation}
Therefore
\begin{equation}
  \label{uve1}
  \fl
  v_1(\mathbf{t}) = \int_J v_1(\mathbf{t}) \rho(x) \rmd x
                           = \int_J V(x) \rho(x) \rmd x + 2 t_0 \int_J\int_{J} \rho(x) \rho(y) \log|x-y| \rmd x \rmd y.
\end{equation}
Moreover, (\ref{0.2}) and~(\ref{1.2}) imply that
\begin{eqnarray}
  \oint_{\gamma} V(z) \rmd S(z) & =  \frac{1}{2} \oint_{\gamma} V(z) h(z) w(z) \rmd z\\
                                                   & = - \int_J V(x) h(x) w_{1+}(x) \rmd x\\
                                                   & = 2\pi\rmi t_0\int_J V(x) \rho(x) \rmd x,
\end{eqnarray}
and we are led to the main result of this section, namely
\begin{eqnarray}
  \fl
  F(\mathbf{t}) = \log\tau(\mathbf{t})
                       \nonumber
                       &= \frac{1}{4 \pi\rmi} \oint_{\gamma} V(z) \rmd S(z) + \frac{t_0}{2} v_{1}\\
                       \label{reee}
                       &= -T \int_{J} V(x) \rho(x) \rmd x + T^2 \int_J\int_J \log|x-y| \rho(x) \rho(y) \rmd x \rmd y,
\end{eqnarray}
where in the last equation we have substituted $t_0$ by $-T$.
\section{Phase transitions in multi-cut matrix models}
For a given fixed potential let us denote by $I_s$  the set of values $T>0$ such that the eigenvalue support $J$ has exactly $s$ cuts and the corresponding eigenvalue density  is regular. These sets correspond to the different \emph{phases} of the matrix model. Common boundary points of two sets $I_s$ and $I_{s-1}$ mark phase transitions in which the number of cuts changes in one unit. Our next goal is to match pairs of solutions of Whitham hierarchies to characterize critical processes.

Hereafter we will use a superscript `$(s)$' to distinguish objects corresponding to different multi-cut cases. Using our new, more explicit notation, we recall that for each $T\in I_s$ the endpoints of the eigenvalue support
\begin{equation}
  \label{c0}
  \beta_1^{(s)}(T)<\beta_2^{(s)}(T)<\ldots<\beta_{2s}^{(s)}(T),
\end{equation}
satisfy the hodograph equations
\begin{equation}
  \label{c1}
  T P_0^{(s)}(\beta_i^{(s)},{\bbeta}^{(s)})
  - \sum_{k=1}^{2p}t_k\,P_k^{(s)}(\beta_i^{(s)},{\bbeta}^{(s)}) = 0,\quad i=1,\ldots,2\,s.
\end{equation}

As we mentioned in the preceding section it is extremely difficult to characterize ${\bbeta}^{(s)}(T)$ from this system. Thus, we will take advantage of the $n=0$ member of the system~(\ref{vel1a})
\begin{equation}
  \label{vel2}
  \dot{\beta}_k^{(s)}
  =
  \frac{4\,P_0^{(s)}(\beta_k^{(s)},{\bbeta}^{(s)})}{h^{(s)}(\beta_k^{(s)},{\bbeta}^{(s)})\,\prod_{i\neq k}(\beta_k^{(s)}-\beta_i^{(s)})},\quad k=1,\ldots,2\,s,
\end{equation}
where the dot denotes derivative with respect to $T$.
\subsection{Phase transitions with $I_s\rightarrow I_{s\pm 1}$}
In this section we will use matched solutions of Whitham hierarchies to describe a type of critical processes in which two consecutive endpoints $(\beta_l^{(s)},\beta_{l+1}^{(s)})$ of the eigenvalue support  coalesce at a critical temperature $T=T_c$, and such that:
\begin{enumerate}
  \item The remaining components of ${\bbeta}^{(s)}$ and  of $\dot{{\bbeta}}^{(s)}$ match continuously with the corresponding limits of ${\bbeta}^{(s-1)}$ and $\dot{{\bbeta}}^{(s-1)}$.
  \item The first and second derivatives $\partial_T F$ and $\partial_{TT} F$ of the free energy are continuous at $T=T_c$.
\end{enumerate}
If we denote by ${\bbeta}^{(s)}_c$ and $\dot{{\bbeta}}^{(s)}_c$ the limits of the endpoints and their $T$-derivatives as $T\rightarrow T_c$, these conditions mean that
\begin{equation}
 \fl
  \label{criti}
  \beta^{(s)}_{c,l}=\beta^{(s)}_{c,l+1},\quad
  (\beta^{(s-1)}_{c,k},\dot{\beta}^{(s-1)}_{c,k})
  = \left\{\begin{array}{l}
    (\beta^{(s)}_{c,k} \dot{\beta}^{(s)}_{c,k}), \quad 1\leq k<l,\\
    \\
    (\beta^{(s)}_{c,k+2},\dot{\beta}^{(s)}_{c,k+2}),\quad l\leq k\leq 2s-2.
    \end{array}\right.
\end{equation}
As an example of this notation, the phase transition in figure~4 corresponds to the case $s=2$, $l=2$ in~(\ref{criti}), where the red labels correspond to the `(2)' phase, the blue labels to the `(1)' phase, $\beta_2^{(2)}$ merges with $\beta_3^{(2)}$, $\beta_1^{(2)}$ matches $\beta_1^{(1)}$, and $\beta_4^{(2)}$ matches $\beta_2^{(1)}$.

As we will see, our method to characterize these processes is based on  the important identity~(\ref{id11}) and the following property of the polynomials $P_k$ (see Theorem 3.5 of~\cite{dl5}): given ${\bbeta}^{(s)}\,(s\geq 2)$ such that for some $l=1,\ldots,2s-1$
\begin{equation}
  \label{cond1}
  \beta_l = \beta_{l+1}= \beta,
\end{equation}
then  the following identities hold
\begin{equation}\label{iw}
  P_k^{(s)}(z,{\bbeta}^{(s)}) = (z-\beta)\,P_k^{(s-1)}(z,{\bbeta}^{(s-1)}),\quad k\geq 0,
\end{equation}
where
\begin{equation}
  \label{cond2}
  \beta_k^{(s-1)} = \left\{\begin{array}{l}
                              \beta^{(s)}_{k},\quad 1\leq k<l,\\
                              \\
                              \beta^{(s)}_{k+2},\quad l\leq k\leq 2s-2.
                              \end{array}\right.
\end{equation}
We give some simple examples of identities~(\ref{iw}) which can be directly verified and that will be used later. 
\subsubsection{Examples}
Consider the polynomials $P_k^{(2)}(z,{\bbeta}^{(2)})$, $(k=0,1)$ with ${\bbeta}^{(2)}=(\beta_1,\beta,\beta,\beta_4)$. From the expressions~(\ref{el1}) and~(\ref{elll}), taking into account that $\beta_2=\beta_3$ implies $r=s=0$, and that~\cite{wol}
\begin{equation}
E(0)=K(0)=\Pi(0,0)=\frac{\pi}{2},
\end{equation}
we deduce at once the identities
\begin{eqnarray}
  P_0^{(2)}&(z,{\bbeta}^{(2)}) = z-\beta = (z-\beta)\,P_0^{(1)}(z,{\bbeta}^{(1)}),\\
  P_1^{(2)}&(z,{\bbeta}^{(2)}) = (z-\beta)\left(\frac{1}{2}\,z-\frac{1}{4}(\beta_1+\beta_4)\right)
                                                         = (z-\beta)\,P_1^{(1)}(z,{\bbeta}^{(1)}).
\end{eqnarray}

Likewise, consider the polynomial $P_0^{(2)}(z,{\bbeta}^{(2)})$ with ${\bbeta}^{(2)}=(\beta_1,\beta_2,\beta,\beta)$. Now $\beta_3=\beta_4$ implies $r=s=1$, and from  the asymptotic approximation~(\ref{bc111}) of Appendix~C it follows that
\begin{equation}
  \lim_{\beta_4\rightarrow \beta_3}\left((\beta_4-\beta_3)\frac{\Pi(r,s)}{K(s)}\right)=0.
\end{equation}
Hence, (\ref{el1}) implies
\begin{equation}
P_0^{(2)}(z,{\bbeta}^{(2)})=z-\beta=(z-\beta)\,P_0^{(1)}(z,{\bbeta}^{(1)}).
\end{equation}
\subsubsection{$C^1$-matching  of eigenvalue supports}
For the critical processes that we are considering the limit values ${\bbeta}_c^{(q)}, (q=s,s-1)$ of the endpoints of the eigenvalue support satisfy the hodograph systems
\begin{equation}\label{c1Q}
  T_c P_{0}^{(q)}(\beta_{c,i}^{(q)},{\bbeta}_c^{(q)})
  -\sum_{k=1}^{2p}t_k\,P_k^{(q)}(\beta_{c,i}^{(q)},{\bbeta}_c^{(q)})
  =0,\quad i=1,\ldots,2q.
\end{equation}
Since ${\bbeta}_c^{(s)}$ verifies the hypothesis~(\ref{cond1}) then from~(\ref{iw}) we deduce that
\begin{equation}
  \label{id02}
  P_k^{(s)}(z,{\bbeta}_c^{(s)})=(z-\beta)\,P_k^{(s-1)} (z,{\bbeta}_c^{(s-1)}).
\end{equation}
As a consequence we have
\begin{equation}
  \label{id2}
  P_k^{(s)}(\beta_{c,i}^{(s)},{\bbeta}_c^{(s)})=(\beta_{c,i}^{(s)}-\beta)\,P_k^{(s-1)}
  (\beta_{c,i}^{(s)},{\bbeta}_c^{(s-1)}),\quad  i=1,\ldots,2s.
\end{equation}
Hence,  the  hodograph equations~(\ref{c1Q}) for $q=s$ and  $i=l,l+1$ are trivially satisfied, while the remaining equations reduce to the system~(\ref{c1Q}) for $q=s-1$. Thus, the hodograph systems are compatible with a continuous matching of the endpoints of the eigenvalue support,
\begin{equation}
  \label{criti02}
  \beta^{(s)}_{c,k} = \beta^{(s-1)}_{c,k'},\quad k'
                             = \left\{\begin{array}{l}
                                         k,\quad 1\leq k<l,\\
                                         \\
                                         k-2,\quad l+2\leq k\leq 2s.
                                        \end{array}\right.
\end{equation}

We notice also that~(\ref{id11}), and~(\ref{id02}) imply
\begin{equation}
  \label{ache}
  h^{(s)}(z,{\bbeta}_c^{(s)}) = \frac{h^{(s-1)}(z,{\bbeta}_c^{(s-1)})}{z-\beta},
\end{equation}
and therefore the continuity of the eigenvalue densities at the critical point follows
\begin{equation}
  \label{cont}
  \rho^{(s)}(x,{\bbeta}_c^{(s)})=\rho^{(s-1)}(x,{\bbeta}_c^{(s-1)}).
  \end{equation}
Furthermore, from~(\ref{vel2}), (\ref{id2}) and~(\ref{ache}) we deduce that
\begin{eqnarray}
  \label{vel1}
  \nonumber &\dot{\beta}_{c,k}^{(s)} =\frac{4\,P_0^{(s)}(\beta_{c,k}^{(s)},{\bbeta}_c^{(s)})}{h^{(s)}(\beta_{c,k}^{(s)},
  {\bbeta}_c^{(s)})\,\prod_{i\neq k}
  (\beta_{c,k}^{(s)}-\beta_{c,i}^{(s)})}\\
 \nonumber \\
  \nonumber &= \frac{4\,P_0^{(s-1)}(\beta_{c,k'}^{(s-1)},{\bbeta}_c^{(s-1)})}{h^{(s-1)}(\beta_{c,k'}^{(s-1)},{\bbeta}_c^{(s-1)})\,\prod_{i\neq k'}
(\beta_{c,k'}^{(s-1)}-\beta_{c,i}^{(s-1)})}\\
\nonumber\\
 & =\dot{\beta}_{c,k'}^{(s-1)},\quad\mbox{$k\neq l,l+1$},
\end{eqnarray}
which shows the $C^1$ property for the matching of the eigenvalue supports.
\subsubsection{Order of the phase transitions}
We will now prove that these critical processes are, generically, third-order phase transitions.

Consider the behaviour of the derivatives of the free energy~(\ref{reee}) with respect to $T$ as $T\rightarrow T_c\pm 0$. From the expression of the free energy~(\ref{reee}) and from~(\ref{cont}) it is clear that  $F$ is continuous at $T_c$. Now~(\ref{idtau}) and~(\ref{multnn}) imply
\begin{equation}
\frac{\partial F(T)}{\partial T}=-v_1=-V(x)+2\,T\,\int_{J}\rho(y)\,\log|x-y|\,\rmd y,\quad x\in J,
\end{equation}
\begin{equation}
  \label{second}
  \frac{\partial^2 F}{\partial T^2}
  = -\partial_T v_1
  = L_0
  = \lim_{Q\rightarrow \infty_1}\left(2\,\log z-\int_{\infty_2}^Q \left(\rmd \Omega_0+\frac{\rmd z}{z-z_0}\right)\right).
\end{equation}
Therefore $\partial_T F$ is also continuous at $T_c$ and from~(\ref{diff1}) and~(\ref{iw}) it follows that
\begin{equation}
  \fl
   \frac{\partial^2 F^{(s)}(T_c)}{\partial T^2}-\frac{\partial^2 F^{(s-1)}(T_c)}{\partial T^2}=\int_{\infty_2}^{\infty_1}
   \left(\frac{P_0^{(s-1)}(z,{\bbeta}_c^{(s-1)})}{w^{(s-1)}(z,{\bbeta}_c^{(s-1)})}-
\frac{P_0^{(s)}(z,{\bbeta}_c^{(s)})}{w^{(s)}(z,{\bbeta}_c^{(s)})}\right) \rmd z=0,
\end{equation}
which shows that $\partial_{TT}F$ is continuous at $T_c$.

Finally, as a consequence of~(\ref{second}),
\begin{equation}
  \frac{\partial^3 F}{\partial T^3}
   = -\int_{\infty_2}^{\infty_1}\,\frac{\partial\rmd \Omega_0}{\partial T}
   = -\int_{\infty_2}^{\infty_1} \,\sum_j\frac{\partial\rmd \Omega_0}{\partial \beta_j}\dot{\beta}_j,
\end{equation}
and using again~(\ref{diff1}) and~(\ref{iw}),
\begin{equation}
\frac{\partial\rmd \Omega^{(s)}_0}{\partial \beta_j^{(s)}}(z,{\bbeta}_c^{(s)})=
\frac{\partial\rmd \Omega^{(s-1)}_0}{\partial \beta_j}(z,{\bbeta}_c^{(s-1)}),\quad j\neq l,l+1.
\end{equation}
Therefore, the jump of the third derivative is
\begin{equation}
  \label{jump}
  \fl
  \frac{\partial^3 F^{(s)}(T_c)}{\partial T^3}-\frac{\partial^3 F^{(s-1)}(T_c)}{\partial T^3}
  = -\int_{\infty_2}^{\infty_1} \,\lim_{T\rightarrow T_c}
     \left(
      \frac{\partial \rmd\Omega^{(s)}_0}{\partial \beta_l^{(s)}}\dot{\beta}_l^{(s)}+
      \frac{\partial \rmd\Omega^{(s)}_0}{\partial \beta_{l+1}^{(s)}}\dot{\beta}_{l+1}^{(s)}\right).
\end{equation}
As we will show with an explicit calculation in section~5, this jump is in general different from zero and, consequently, these processes represent third-order phase transitions.
\section{Phase transitions $I_1\leftrightarrow I_2$}
In this section we first particularize our previous analysis to the case in which $T_c$ is a common boundary point of two regions $I_1$ and $I_2$ and then apply these results to the matrix model corresponding to the Bleher-Eynard potential~\cite{be}.

For $s=1$, (\ref{q1}) implies that the hodograph equations~(\ref{c1}) can be written in the form
\begin{equation}
  \label{hod1}
  \left.\left(\left(
    \frac{2T}{z} - V'(z)\right) w_1^{(1)}(z,{\bbeta}^{(1)})\right)_\oplus\right|_{z=\beta_i^{(1)}}=0,
  \quad i=1,2,
\end{equation}
where
\begin{equation}
  w_1^{(1)}(z,{\bbeta}^{(1)}) = \sqrt{(z-\beta_1^{(1)}) (z-\beta_2^{(1)})}.
\end{equation}
Moreover, (\ref{vel2}) reduces to
\begin{eqnarray}
  \label{vs1}
  \dot{\beta}_1^{(1)} = \frac{4}{h^{(1)}(\beta_1^{(1)},{\bbeta}^{(1)})
                                                    \left(\beta_1^{(1)}-\beta_2^{(1)}\right)},\\
  \label{vs2}
  \dot{\beta}_2^{(1)} = \frac{4}{h^{(1)}(\beta_2^{(1)},{\bbeta}^{(1)})
                                                    \left(\beta_2^{(1)}-\beta_1^{(1)}\right)} .
\end{eqnarray}
Similarly, from~(\ref{q2}) it follows that the hodograph equations~(\ref{c1}) for $s=2$ can be written
\begin{equation}
  \label{hod2}
  \fl
  \left.\left(\left(
    \frac{2T}{z} - V'(z)\right) w_1^{(2)}(z,{\bbeta}^{(2)})\right)_\oplus\right|_{z=\beta_i^{(1)}}
    + U(T,{\bbeta}^{(2)})=0,\quad i=1,2,3,4,
\end{equation}
where
\begin{equation}
  w_1^{(2)}(z,{\bbeta}^{(2)})
  =
  \sqrt{(z-\beta_1^{(2)}) (z-\beta_2^{(2)}) (z-\beta_3^{(2)}) (z-\beta_4^{(2)})},
\end{equation}
and
\begin{eqnarray}
  \nonumber
  \fl
  U(T,{\bbeta}^{(2)})
  =
  - \frac{\sqrt{(\beta^{(2)}_4-\beta^{(2)}_2)(\beta^{(2)}_3-\beta^{(2)}_1)}}{2\,K(s({\bbeta}^{(2)}))}
  \times\\
  \int_{\beta^{(2)}_2}^{\beta^{(2)}_3}
        \left(\left(\frac{2T}{x}-V'(x)\right)w_1^{(2)}(x,{\bbeta}^{(2)})\right)_{\oplus}
        \frac{\rmd x}{w_1^{(2)}(x,{\bbeta}^{(2)})}.
\end{eqnarray}
Although these hodograph equations are extremely involved, the corresponding ordinary differential equations~(\ref{vel2}) for $\bbeta^{(2)}$ are much simpler. Indeed:
\begin{equation}
  \label{merr2}
  \dot{\beta}_i^{(2)}
  =
  \frac{4(\beta_i^{(2)}-C({\bbeta}^{(2)}))}{
           h^{(2)}(\beta_i^{(2)},{\bbeta}^{(2)})\,\prod_{k\neq i}(\beta_i^{(2)}-\beta_k^{(2)})},
           \quad i=1,\ldots,4,
\end{equation}
where
\begin{equation}
  r = r({\bbeta}^{(2)}) = \frac{\beta_3^{(2)}-\beta_2^{(2)}}{\beta_4^{(2)}-\beta_2^{(2)}},
\end{equation}
\begin{equation}
  s = s({\bbeta}^{(2)} )=
  \frac{(\beta_4^{(2)}-\beta_1^{(2)})(\beta_3^{(2)}-\beta_2^{(2)})}{
           (\beta_3^{(2)}-\beta_1^{(2)})(\beta_4^{(2)}-\beta_2^{(2)})},
\end{equation}
and
\begin{equation}
   C({\bbeta}^{(2)}) = \beta_4^{(2)}-(\beta_4^{(2)}-\beta_3^{(2)})\,\frac{\Pi(r,s)}{K(s)}.
\end{equation}

As we have seen in the preceding subsection the second derivative $\partial_{TT}F$ of the free energy (the specific heat)
\begin{equation}
  \frac{\partial^2 F}{\partial T^2}
  = \lim_{Q\rightarrow\infty_1} \left(2 \log z - \int_{\infty_2}^Q \left(\rmd \Omega_0+\frac{\rmd z}{z-z_0}\right)\right),
\end{equation}
 is continuous at $T=T_c$. Then in  a transition $I_1\leftrightarrow I_2$ it can be calculated from the limit value of $\partial_{TT} F^{(1)}$. In this way, integrating from $\infty_2$ to $\infty_1$ along to the path shown in figure~5  and taking into account that 
\begin{equation}
  \rmd \Omega_0^{(1)} = \frac{\rmd z}{\sqrt{(z-\beta_1^{(1)})(z-\beta_2^{(1)})}},
\end{equation}
we get   
\begin{equation}
  \frac{\partial^2 F(T_c)}{\partial T^2} = 2 \log\frac{ \beta_2^{(1)} - \beta_1^{(1)} }{4}.
\end{equation}
\begin{figure}
  \includegraphics[width=6cm]{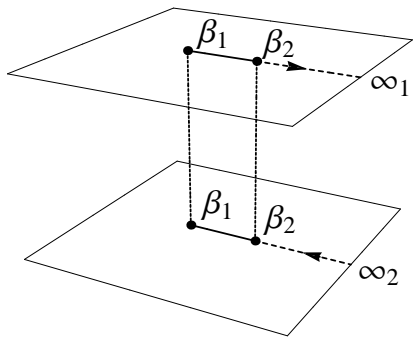}
  \caption{Integration path from $\infty_2$ to $\infty_1$.}
  \label{fig:ip}
\end{figure}
\subsection{Merging and birth of cuts in the Bleher-Eynard model}
The Bleher-Eynard potential~\cite{be}
\begin{equation}
  V(x) = \frac{1}{4}\,x^4-\frac{4}{3}\,c\,x^3+(2\,c^2-1)\,x^2+8\,c\,x,\quad -1<c<1,
  \label{eq:bepot}
\end{equation}
defines a  matrix model exhibiting interesting critical processes. According to the upper bound~(\ref{nice}) there are two possible phases $s=1,2$ only. In these cases it follows easily that
\begin{eqnarray}
  \fl
  \nonumber
  h^{(1)}(z,{\bbeta}^{(1)})
  = \left(z - 2 c+\frac{\beta_1^{(1)}+\beta_2^{(1)}}{4}\right)^2
    + \frac{5}{16}\left(\beta_1^{(1)}+\beta_2^{(1)}\right)^2\\
    {}- \frac{1}{2}\,\beta_1^{(1)}\,\beta_2^{(1)}-c\,(\beta_1^{(1)}+\beta_2^{(1)})-2,
  \label{h1}
\end{eqnarray}
and
\begin{equation}
  \label{h2}
  h^{(2)}(z,{\bbeta}^{(2)})
  =
  z + \frac{1}{2} (\beta_1^{(2)}+\beta_2^{(2)}+\beta_3^{(2)}+\beta_4^{(2)}) - 4 c.
\end{equation}
\subsubsection*{Merging of two cuts}
We first look for a particular solution of the hodograph equations. A direct calculation shows that
\begin{equation}
  \fl
  \left(\left(\frac{2T}{z}-V'(z)\right) w_1^{(1)}(z,{\bbeta}^{(1)})\right)_{\oplus}=
  2 T - z^4 + 4 c z (z^2-4) - 4 (c^2-1) z^2 + 8 c^2 - 2,
\end{equation}
and as a consequence it is easy to see that at $T_c=1+4 c^2$ the hodograph equations~(\ref{hod1}) are satisfied by ${\bbeta}_c^{(1)}=(\beta_{c,1}^{(1)},\beta_{c,2}^{(1)})=(-2,2)$. Moreover, we have
\begin{equation}
  h^{(1)}(z,{\bbeta}_c^{(1)}) = (z-2c)^2,
\end{equation}
so that the eigenvalue density vanishes at $x=2 c\in J$. Hence $T_c$ is a  a common boundary point
of  $I_1$ and $I_{2}$, and  the endpoints $(\beta_2^{(2)},\beta_{3}^{(2)})$ of the eigenvalue support for $s=2$ tend to $\beta=2\,c$ as $T\rightarrow T_c$ in $I_2$. Thus
\begin{equation}
  \beta_{c,1}^{(2)} = -2, \quad
  \beta_{c,2}^{(2)} = \beta_{c,3}^{(2)} = \beta,\quad
  \beta_{c,4}^{(2)}=2,
\end{equation}
and 
\begin{equation}
  r({\bbeta}_c^{(2)}) = s({\bbeta}_c^{(2)}) = 0,\quad
  h^{(2)}(z,{\bbeta}_c^{(2)}) = z-\beta.
\end{equation}
with $-2<\beta=2\,c<2$.

From~(\ref{vs1}) we obtain
\begin{equation}
  \label{vel2a}
  \dot{\beta}_{c,1}^{(2)} = \dot{\beta}_{c,1}^{(1)}
                                         = \frac{-1}{(\beta+2)^2},\quad
  \dot{\beta}_{c,4}^{(2)} = \dot{\beta}_{c,2}^{(1)}
                                         = \frac{1}{(\beta-2)^2}.
\end{equation}
If we write
\begin{equation}
  h^{(1)}(z,\bbeta^{(1)})
 = \left(z-2c+\frac{\beta_1^{(1)}+\beta_2^{(1)}}{4}\right)^2
   + \alpha(\bbeta^{(1)}),
\end{equation}
where
\begin{equation}
  \alpha(\bbeta^{(1)})
 =
 \frac{5}{16}\left(\beta_1^{(1)}+\beta_2^{(1)}\right)^2
 - \frac{1}{2}\,\beta_1^{(1)}\,\beta_2^{(1)}-c\,(\beta_1^{(1)}+\beta_2^{(1)})-2,
\end{equation}
and use the expressions~(\ref{vel2a}) for the velocities $\dot{\beta}_{c,i}^{(1)}\, (i=1,2)$ we get
\begin{equation}
  \left.\frac{\rmd}{\rmd T}\alpha(\bbeta^{(1)}(T))\right|_{T=T_c}=\frac{1}{2(1-c^2)}>0.
\end{equation}
This implies that for $T$ slightly to the right of $T_c$ the function $h^{(1)}(x,\bbeta^{(1)}(T))$ is strictly positive for all real $x$ and $T$ is in $I_1$. Consequently, if $T$ slightly to the left of $T_c$ then $T$ is in $I_2$. Therefore, the critical process at $T_c$ represents a merging of two cuts.

From the expressions~(\ref{vel2a}) for the velocities  $\dot{\beta}_{c,i}^{(2)}$ it follows that as $T\rightarrow T_c-0$
\begin{equation}
  \label{unocuatro}
  \beta_1^{(2)}(T_c-t) \sim -2+\frac{t}{(\beta+2)^2},\quad
  \beta_4^{(2)}(T_c-t) \sim 2-\frac{t}{(\beta-2)^2},
\end{equation}
whereas the expressions~(\ref{merr2}) for $\dot{\beta}_2^{(2)}$ and $\dot{\beta}_3^{(2)}$
become undetermined because  $\beta_i^{(2)}-C({\bbeta}^{(2)})$, $h(\beta_i^{(2)},{\bbeta})$ and $\beta_2^{(2)}-\beta_3^{(2)}$ vanish. However, we show in Appendix~C (equations~(\ref{cees})--(\ref{e6})) that
\begin{equation}
  C({\bbeta}^{(2)})\sim \beta_2^{(2)}+\frac{1}{2}\,(\beta_3^{(2)}-\beta_2^{(2)}),
\end{equation}
and therefore~(\ref{merr2}) leads to
\begin{eqnarray}
  \dot{\beta}_2^{(2)} \sim
  \frac{2}{h(\beta_2^{(2)},{\bbeta}^{(2)})(\beta_2^{(2)}-\beta_1^{(2)})(\beta_2^{(2)}-\beta_4^{(2)})},
  \\
  \dot{\beta}_3^{(2)} \sim
  \frac{2}{h(\beta_3^{(2)},{\bbeta}^{(2)})(\beta_3^{(2)}-\beta_1^{(2)})(\beta_3^{(2)}-\beta_4^{(2)})},
\end{eqnarray}
which imply  the following asymptotic behaviours~\cite{be}: 
\begin{equation}
  \label{dostres}
  \beta_2^{(2)}(T_c-t) \sim \beta-\frac{2 t^{1/2}}{\sqrt{4-\beta^2}},\quad
  \beta_3^{(2)}(T_c-t)\sim \beta+\frac{2 t^{1/2}}{\sqrt{4-\beta^2}}.
\end{equation}
\subsubsection*{Analysis of the third-order phase transition at $T=T_c$}
We have that
\begin{equation}
  \label{jump1}
  \frac{\partial^3 F}{\partial T^3}
  =
  -\int_{\infty_2}^{\infty_1}\frac{\partial\rmd \Omega_0}{\partial T}
  =
  -\int_{\infty_2}^{\infty_1}\,\frac{\partial}{\partial T}\left( \frac{P_0(z,{\bbeta})}{w(z,{\bbeta})}\right)\rmd z,
\end{equation}
and it is clear that
\begin{equation}
  \nonumber
  \fl
  \frac{P_0^{(1)}(z,{\bbeta}^{(1)})}{w^{(1)}(z,{\bbeta}^{(1)})}
  -
  \frac{P_0^{(2)}(z,{\bbeta}^{(2)})}{w^{(2)}(z,{\bbeta}^{(2)})}
  \sim
  \frac{\sqrt{(z-\beta_2^{(2)})(z-\beta_3^{(2)})}-(z-C({\bbeta}^{(2)}))}{(z-\beta)\sqrt{z^2-4}}.
\end{equation}
If we now use the Taylor expansion at ${\bbeta}_r^{(2)}=(\beta_1^{(2)},\beta_2^{(2)},\beta_2^{(2)},\beta_4^{(2)})$ 
\begin{eqnarray}
  \nonumber
  \fl
  \sqrt{(z-\beta_2^{(2)})(z-\beta_3^{(2)})}
   &= (z-\beta_2)
        \sqrt{1-\frac{\beta_3^{(2)}-\beta_2^{(2)}}{z-\beta_2^{(2)}}}\\
    \fl
    &= z-\beta_2
         -\frac{\beta_3^{(2)}-\beta_2^{(2)}}{2}
         -\frac{(\beta_3^{(2)}-\beta_2^{(2)})^2}{8(z-\beta_2^{(2)})}+
         \mathcal{O}((\beta_3^{(2)}-\beta_2^{(2)})^3),
\end{eqnarray}
then, taking into account~(\ref{tay}), we have that as $T\rightarrow T_c$
\begin{equation}
  \label{a22}
  \fl
  \sqrt{(z-\beta_2^{(2)})(z-\beta_3^{(2)})} - (z-C({\bbeta}^{(2)})
  \sim
  \frac{1}{8}\left(-\frac{1}{z-\beta}+\frac{\beta}{4-\beta^2}\right)(\beta_3-\beta_2)^2.
\end{equation} Hence
it follows from~(\ref{dostres}) that
\begin{equation}
  \label{sss}
  \fl
  \frac{P_0^{(1)}(z,{\bbeta}^{(1)})}{w^{(1)}(z,{\bbeta}^{(1)})}
  -
  \frac{P_0^{(2)}(z,{\bbeta}^{(2)})}{w^{(2)}(z,{\bbeta}^{(2)})}
  \sim\
  \frac{2(T-T_c)}{(4-\beta^2)\sqrt{z^2-4}}
  \left(\frac{1}{(z-\beta)^2}-\frac{\beta}{(4-\beta^2)(z-\beta)}\right).
\end{equation}
Therefore integrating from $\infty_2$ to $\infty_1$ along the path shown in figure~5 we get
\begin{eqnarray}
  \label{ccut}
  \nonumber
  \fl
  \frac{\partial^3 F^{(2)}(T_c)}{\partial T^3}-\frac{\partial^3 F^{(1)}(T_c)}{\partial T^3}
  \\
  \nonumber
  = \frac{4}{4-\beta^2}
  \left( \int_2^{\infty} \frac{\rmd x}{(x-\beta)^2\,\sqrt{x^2-4}}-
  \frac{\beta}{4-\beta^2}\,\int_2^{\infty} \frac{\rmd x}{(x-\beta)\,\sqrt{x^2-4}}\right)\\
  = \frac{4}{(4-\beta^2)^{2}}.
\end{eqnarray}
This equation shows that the third-order derivative of the free energy jumps at the critical point~\cite{be}. 
\subsubsection*{Numerical calculations}
In section~3.1.1 we have solved directly the simple hodograph equations~(\ref{eq:hods1a})--(\ref{eq:hods1b}) for the symmetric quartic potential~(\ref{eq:bepotc0}) in the $s=1$ region and, by imposing in advance the symmetry relations between the roots, we have solved also the hodograph equations~(\ref{eq:hods2a})--(\ref{eq:hods2b}) for the same potential in the $s=2$ region. The symmetric quartic potential~(\ref{eq:bepotc0}) is the particular case $c=0$ of the Bleher-Eynard potential~(\ref{eq:bepot}), but nonsymmetric (i.e., $c\neq 0$) cases of this latter potential are not amenable to the same direct procedure. The corresponding hodograph equations in the $s=1$ region are still purely algebraic, and with the help of a symbolic manipulation program the solutions can be found implicitly among the roots a two-variable polynomial equation of degree 12. The hodograph equations in the $s=2$ region are already nonalgebraic (the roots appear in the arguments of elliptic functions) and even a numerical solution and identification of the roots seems to be extremely difficult. However, we have shown that the roots also satisfy the system of first order ordinary differential equations~(\ref{vel2}), which is readily amenable to numerical integration. For concreteness we will outline the procedure in the particular case $c=1/2$, which turns out to be typical, and omit the superscripts (cf. figure~6).

We take as our starting point the critical temperature $T_c=2$ (marked by the rightmost dashed vertical line in figure~6), for which we have already discussed the solution of the hodograph equations. We have also shown that sufficiently small values of $T$ to the right of $T_c$ belong to an $s=1$ interval, and that at $T=T_c$ the system~(\ref{vel2}) for $s=1$ is regular (the denominators do not vanish). Therefore for $T>T_c$ we can integrate the system~(\ref{vel2}) directly with the initial condition $\beta_1(T_c)=-2$, $\beta_2(T_c)=2$. In this interval the system is numerically well behaved and we obtain the two curves shown in figure~6. We have checked that within our numerical precision these solutions coincide with two arcs of the algebraic curve mentioned in the previous paragraph. Note also that the mere success of the numerical integration does not guarantee that the system remains in an $s=1$ region and we have to study later the behavior of the density of states.

Likewise, sufficiently small values of $T$ to the left of $T_c$ belong to an $s=2$ interval, but at  $T=T_c$ the system~(\ref{vel2}) for $s=2$ is singular and we cannot start our integration to the left simply with the initial condition $\beta_1(T_c)=-2$, $\beta_2(T_c)=\beta_3(T_c)=1$, $\beta_4(T_c)=2$. We circumvent this problem by taking as our initial point a value $T_0=T_c-t$ with $t$ sufficiently small, and using the asymptotic behaviors~(\ref{unocuatro}) and~(\ref{dostres}) as initial values. As the integration proceeds to the left we find that the solutions $\beta_3(T)$ and $\beta_4(T)$ coalesce at a second critical temperature $\widetilde{T}_c\approx 1.845\,097$, where we find numerically a birth of a cut. This second critical temperature is marked by the leftmost vertical dashed line in figure~6. Incidentally, the computational effort required by the adaptive numerical integrator increases as the integration approaches the $\widetilde{T}_c$, where the system of differential equations is again singular and the arguments of the ensuing elliptic integrals tend to one. We have found that we can achieve higher precision if at an intermediate point, say $T=1.85$, we rewrite the system of differential equations using the relations~\cite{wol2,wol3}
\begin{equation}
  \fl
  K(s) = \frac{1}{\sqrt{1-s}}K\left(\frac{s}{s-1}\right),\quad |\arg(1-s)|<\pi,
\end{equation}
\begin{equation}
  \fl
  \Pi(r,s) = \frac{1}{(1-r)\sqrt{1-s}}\Pi\left(\frac{r}{r-1},\frac{s}{s-1}\right),\quad |\arg(1-r)|<\pi, |\arg(1-s)|<\pi,
\end{equation}
to map the branch points of the elliptic integrals at $1$ to $-\infty$.

Finally, from $\widetilde{T}_c$ to $T=0$ we integrate again the nonsingular system of differential equations~(\ref{vel2}) for $s=1$ taking as initial conditions the values we have obtained in the previous step for $\beta_1(\widetilde{T}_c)$ and $\beta_2(\widetilde{T}_c)$, and get the two curves shown in figure~6 that coalesce at $T=0$. As a double check of this last step we point out that, within our numerical precision, the solutions again coincide with two arcs of the of the degree 12 algebraic curve mentioned previously. Moreover, the point $\beta_1(0)=\beta_2(0)$ can be calculated exactly, because it is the position  $\beta_{{\min}}$ of the  minimum of the potential $V(x)$, i.e. the real root of the cubic equation
\begin{equation}
  \beta_{{\min}}^3 - 4 c \beta_{{\min}}^2 + 2 (2c^2-1) \beta_{{\min}} + 8 c=0,
\end{equation}
or, using the Cardano formula,
\begin{equation}
\fl
 \beta_{{\min}}=\frac{4}{3} c+\frac{2^{1/3}}{3}\left(-4 c^3-36 c+3 \sqrt{6} \sqrt{\Delta}\right)^{1/3}+\frac{4 c^2+6}{2^{1/3}3
\left(-4 c^3-36 c+3 \sqrt{6} \sqrt{\Delta}\right)^{1/3}},
\end{equation}
where $\Delta=4\,c^4+22\,c^2-1$.  In particular, for our case $c=1/2$ we find
\begin{equation}
  \beta_{{\min}} = \frac{1}{3}
   \left(2-\frac{7}{\sqrt[3]{3
   7-3
   \sqrt{114}}}-\sqrt[3]{37-3
   \sqrt{114}}\right)
   \approx -1.269\,53,
\end{equation}
in agreement with our numerical results.
\begin{figure}
  \includegraphics[width=10cm]{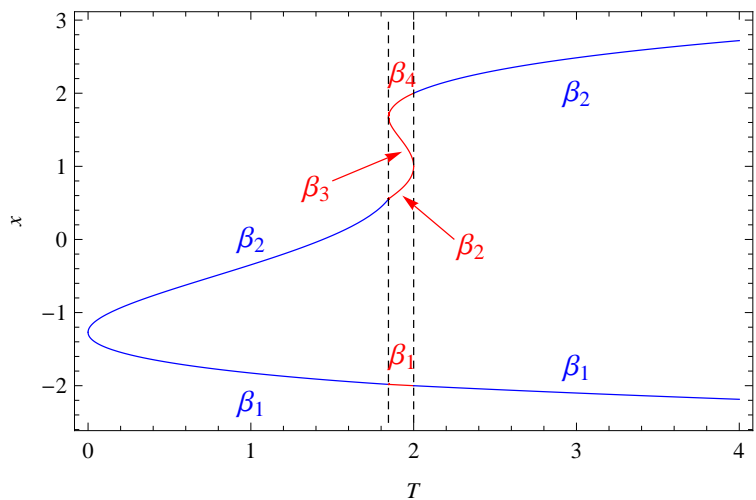}
  \caption{Evolution of the endpoints for the Bleher-Eynard potential with $c=1/2$.}
\end{figure}

We next use~(\ref{0.2}) and~(\ref{0.3}) to calculate the density $\rho(x)$ in each of the three subintervals $0<T<\widetilde{T}_c$ ($s=1$), $\widetilde{T}_c<T<T_c$ ($s=2$), and $T_c<T<T_{\max}$ ($s=1$), and verify that $\rho(x)>0$ for all $x\in J$ and that $\rho(x)$ is normalized. As a sample of our numerical results, in figure~7 we plot the density as a function of the position $x$ for the first critical temperature $\widetilde{T}_c\approx 1.845\,097$, for a temperature $T=1.9$ in the $s=2$ interval, for the second critical temperature $T_c=2$ and for a higher temperature in the second $s=1$ interval $T=3$. Within our numerical precision all these densities are normalized. The transition from the first to the second plot illustrates clearly the birth of a cut; as the temperature increases both intervals lengthen until at $T_c=2$ they join in the third plot, illustrating the merging of the two cuts; finally, in the last plot we see the normalized, one interval density after the two cuts have merged.
\begin{figure}
  \includegraphics[width=12cm]{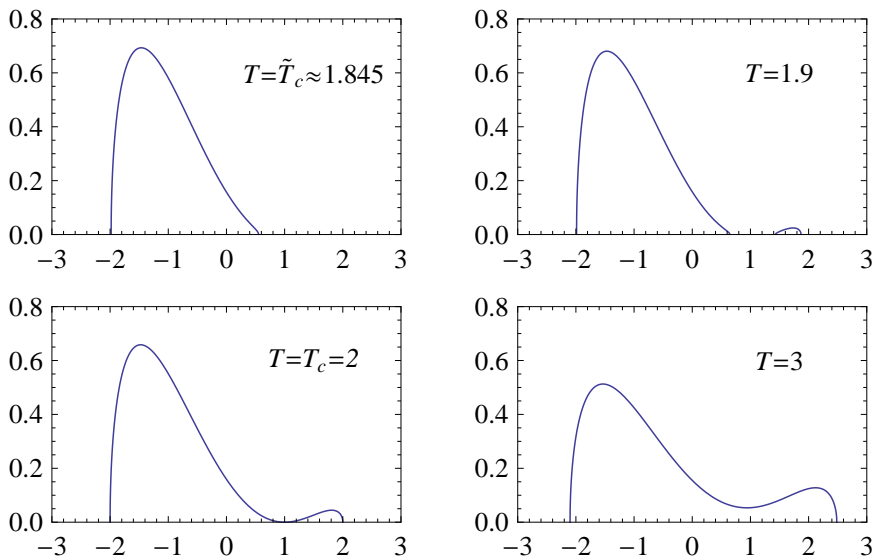}
  \caption{Density $\rho(x)$ as a function of the position $x$ for the Bleher-Eynard potential with $c=1/2$ at four temperatures. The first two plots illustrate the birth of a cut at $\widetilde{T}_c\approx 1.845\,097$; the last two plots illustrate the merging of two cuts at $T_c=2$.}
\end{figure}

Finally, in figure~8 we illustrate the sufficient conditions~(\ref{des1}), (\ref{des2}) and~(\ref{des3}). The figure corresponds to the second plot of figure~7, i.e., to a temperature $T=1.9$, for which the solutions are $\beta_1\approx -1.989$,  $\beta_2\approx 0.646$,  $\beta_3\approx 1.431$ and  $\beta_4\approx 1.870$. Since this temperature belongs to the $s=2$ interval we have the three conditions,
\begin{eqnarray}
  \label{eq:s1}
  &\int_x^{\beta_1}     h(x') w_{1}(x') \rmd x'\leq 0,\quad \mbox{for $x<\beta_1$},\\
  \label{eq:s2}
  &\int_{\beta_{2}}^x  h(x') w_{1}(x') \rmd x'\geq 0,\quad \mbox{for $\beta_{2}<x<\beta_3$},\\
  \label{eq:s3}
  &\int_{\beta_{4}}^x h(x') w_{1}(x') \rmd x'\geq 0,\quad \mbox{for $x>\beta_4$},
\end{eqnarray}
which correspond respectively to the three plots.
\begin{figure}
  \includegraphics[width=15cm]{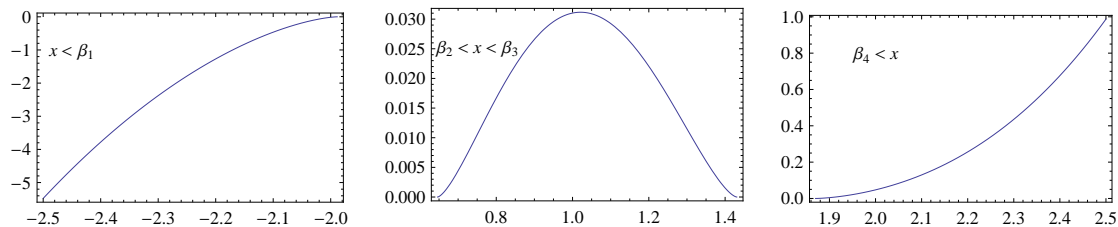}
  \caption{Illustration of the sufficient conditions~(\ref{eq:s1}), (\ref{eq:s2}) and~(\ref{eq:s3}) for the Bleher-Eynard potential with $c=1/2$ in the $s=2$ region ($T=1.9$). The approximate values of the endpoints are $\beta_1\approx -1.989$,  $\beta_2\approx 0.646$,  $\beta_3\approx 1.431$ and  $\beta_4\approx 1.870$, and the plots show the respective integrals as a function of $x$.}
\end{figure}
\subsubsection*{Birth of a cut}
We have seen that temperatures slightly below $T_c$ belong to a two-cut region $I_2$. In the particular case $c=0$ where Bleher-Eynard potential reduces to the quartic even potential studied in 3.1.1,  in which this two-cut region is the whole interval $0<T<T_c$. However, for  $c=1/2$ the numerical integration of the differential equations~(\ref{vel2}) has revealed a birth of a cut at a certain temperature $\widetilde{T}_c$ between $0$ and $T_c$. According to the variational equations~(\ref{2a}) and~(\ref{2b}), as $T\rightarrow 0$ the closure of the eigenvalue support reduces to the set of absolute minima of the potential $V(x)$. But, except for $c=0$, this set is  given by the single point  $\beta_{\min}$, and then a birth of a cut must happen at a certain $0<\widetilde{T}_c< T_c$ for all $c\neq 0$.

We denote the endpoints of the support at the critical temperature $T=\widetilde{T}_c$ by
$\widetilde{\beta}_{c,1}$, $\widetilde{\beta}_{c,2}$, and $\widetilde{\beta}_{c,3}^{(2)} = \widetilde{\beta}_{c,4}^{(2)} = \widetilde{\beta}
$. Note that
\begin{equation}
  r(\widetilde{{\bbeta}}_c^{(2)})
  =
  s(\widetilde{{\bbeta}}_c^{(2)})=1,
\end{equation}
and
\begin{equation}
  h^{(2)}(z,\widetilde{{\bbeta}}_c^{(2)})
  = z + \frac{1}{2}\,(\widetilde{\beta}_{c,1}^{(2)}+\widetilde{\beta}_{c,2}^{(2)}) + \widetilde{\beta} -4 c.
\end{equation}
We prove in Appendix~C that as $T\rightarrow\widetilde{T}_c$
\begin{equation}
  \label{eq:piovk}
  \frac{\Pi(r,s)}{K(s)} \sim
  -\frac{2\sqrt{(\widetilde{\beta}-\widetilde{\beta}_{c,1})(\widetilde{\beta}-\widetilde{\beta}_{c,2})}}{
    \left(\beta_4^{(2)}-\beta_3^{(2)}\right)
    \log(\beta_4^{(2)}-\beta_3^{(2)})}
    \tanh^{-1}\left(\sqrt{\frac{\widetilde{\beta}-\widetilde{\beta}_{c,2}}{\widetilde{\beta}-\widetilde{\beta}_{c,1}}}\right).
\end{equation}
As a consequence,
\begin{equation}
  C(\widetilde{{\bbeta}}_c^{(2)}) \sim \widetilde{\beta}_{4}^{(2)}.
\end{equation}
Thus~(\ref{vs1}) leads to
\begin{eqnarray}
  \label{vel21}
  \beta_1^{(2)}(\widetilde{T}_c+t) \sim \widetilde{\beta}_{c,1}^{(2)} +
  \frac{4 t}{h^{(2)}(\widetilde{\beta}_{c,1}^{(2)},\widetilde{{\bbeta}}_c^{(2)})
                  \left(\widetilde{\beta}_{c,1}^{(2)}-\widetilde{\beta}_{c,2}^{(2)}\right)
                  \left(\widetilde{\beta}_{c,1}^{(2)}-\widetilde{\beta}\right)},
  \\
  \label{vel22}
  \beta_2^{(2)}(\widetilde{T}_c+t) \sim \widetilde{\beta}_{c,2}^{(2)} +
  \frac{4\,t}{h^{(2)}(\widetilde{\beta}_{c,2}^{(2)},\widetilde{{\bbeta}}_c^{(2)})
                   \left(\widetilde{\beta}_{c,2}^{(2)}-\widetilde{\beta}_{c,1}^{(2)}\right)
                   \left(\widetilde{\beta}_{c,2}^{(2)}-\widetilde{\beta}\right)}.
\end{eqnarray}

In order to characterize the behaviour of $\beta_3^{(2)}$ and $\beta_4^{(2)}$ we use~(\ref{merr2}),
which now implies
\begin{eqnarray}
  \label{betasd1}
  \dot{\beta}_3^{(2)} = \frac{4(1-\Pi(r,s)/K(s))}{h^{(2)}(\beta_3^{(2)},{\bbeta}^{(2)})
                                                \left(\beta_3^{(2)}-\beta_1^{(2)}\right)
                                                \left(\beta_3^{(2)}-\beta_2^{(2)}\right)},\\
  \label{betasd2}
  \dot{\beta}_4^{(2)} = \frac{4\,\Pi(r,s)/K(s)}{h^{(2)}(\beta_4^{(2)},{\bbeta}^{(2)})
                                                \left(\beta_4^{(2)}-\beta_1^{(2)}\right)
                                                \left(\beta_4^{(2)}-\beta_2^{(2)}\right)}.
\end{eqnarray}
Using~(\ref{eq:piovk}) it is easy to see that~(\ref{betasd1}) and~(\ref{betasd2}) imply that
\begin{eqnarray}
  \label{betasdd1}
  (\beta_4^{(2)}-\beta_3^{(2)}) \log (\beta_4^{(2)}-\beta_3^{(2)}) \dot{\beta}_3^{(2)} \sim \gamma,\\
  \label{betasdd2}
  (\beta_4^{(2)}-\beta_3^{(2)}) \log (\beta_4^{(2)}-\beta_3^{(2)})\,\dot{\beta}_4^{(2)}\sim -\gamma,
\end{eqnarray}
where
\begin{equation}
  \gamma
  =
  \frac{16\tanh^{-1}\left(\sqrt{\frac{\widetilde{\beta}-\widetilde{\beta}_{c,2}}{\widetilde{\beta}-\widetilde{\beta}_{c,1}}}\right)
          }{
          \left(4\widetilde{\beta}+\widetilde{\beta}_{c,1}+\widetilde{\beta}_{c,2}-8c\right)
          \sqrt{(\widetilde{\beta}-\widetilde{\beta}_{c,1})(\widetilde{\beta}-\widetilde{\beta}_{c,2})}}.
\end{equation}
In this way, solving~(\ref{betasdd1}) and~(\ref{betasdd2}) yields~\cite{ey2}
\begin{eqnarray}
  \label{bc1}
  \beta_3^{(2)}(\widetilde{T}_c+t) \sim \widetilde{\beta} - \left(-2\gamma\frac{t}{\log t}\right)^{1/2},\\
  \label{bc2}
  \beta_4^{(2)}(\widetilde{T}_c+t) \sim \widetilde{\beta} + \left(-2\gamma\frac{t}{\log t}\right)^{1/2}.
\end{eqnarray}
\subsubsection*{Analysis of the third-order phase transition at $T=\widetilde{T}_c$}
In this case the difference between the third-order derivatives of the free energy above and below the critical temperature diverges~\cite{ey2}.  Indeed, proceeding as in the analysis of the merging of two cuts we have 
\begin{equation}
  \label{jump11}
  \frac{\partial^3 F}{\partial T^3}
  = -\int_{\infty_2}^{\infty_1}\frac{\partial\rmd \Omega_0}{\partial T}
  = -\int_{\infty_2}^{\infty_1}\frac{\partial}{\partial T}
                                               \left(\frac{P_0(z,{\bbeta})}{w(z,{\bbeta})}\right)\rmd z,
\end{equation}
\begin{eqnarray}
  \nonumber
  \fl
  \frac{P_0^{(1)}(z,{\bbeta}^{(1)})}{w^{(1)}(z,{\bbeta}^{(1)})}
  -
  \frac{P_0^{(2)}(z,{\bbeta}^{(2)})}{w^{(2)}(z,{\bbeta}^{(2)})}
  \sim
  \frac{1}{(z-\widetilde{\beta})\sqrt{(z-\widetilde{\beta}_{c,1})(z-\widetilde{\beta}_{c,2})}}\\
  \label{A22}
  {}\times\left(\sqrt{(z-\beta_3^{(2)})(z-\beta_4^{(2)})}-(z-C({\bbeta}^{(2)})\right).
\end{eqnarray}
We now use the following Taylor expansion at ${\bbeta}_r^{(2)}=(\beta_1^{(2)},\beta_2^{(2)},\beta_3^{(2)},\beta_3^{(2)})$:
\begin{eqnarray}
  \nonumber
  \sqrt{(z-\beta_3^{(2)})(z-\beta_4^{(2)})}
  &= (z-\beta_3^{(2)})\sqrt{1-\frac{\beta_4^{(2)}-\beta_3^{(2)}}{z-\beta_3^{(2)}}}\\
  &= z-\beta_3-\frac{\beta_4^{(2)}-\beta_3^{(2)}}{2}+\mathcal{O}((\beta_4^{(2)}-\beta_3^{(2)})^2).
\end{eqnarray}
Furthermore, (\ref{eq:piovk}) implies that 
\begin{equation}
  C({\bbeta}^{(2)})
  \sim
  \widetilde{\beta}
  +
  \frac{2\sqrt{(\widetilde{\beta}-\widetilde{\beta}_{c,1})(\widetilde{\beta}-\widetilde{\beta}_{c,2})}
          }{
          \log(\beta_4^{(2)}-\beta_3^{(2)})}
  \tanh^{-1}\left(\sqrt{\frac{\widetilde{\beta}-\widetilde{\beta}_{c,2}}{\widetilde{\beta}-\widetilde{\beta}_{c,1}}}\right).
\end{equation}
Hence, we deduce  that 
\begin{equation}
  \label{A2}
  \frac{P_0^{(1)}(z,{\bbeta}^{(1)})}{w^{(1)}(z,{\bbeta}^{(1)})}
  -
  \frac{P_0^{(2)}(z,{\bbeta}^{(2)})}{w^{(2)}(z,{\bbeta}^{(2)})}
  \sim
  \frac{\widetilde{\gamma}}{\log (T-\widetilde{T}_c)},
\end{equation}
where
\begin{equation}
  \widetilde{\gamma}
  =
  \frac{4\sqrt{(\widetilde{\beta}-\widetilde{\beta}_{c,1})(\widetilde{\beta}-\widetilde{\beta}_{c,2})}
          }{
          \left(z-\widetilde{\beta}\right)
          \sqrt{(z-\widetilde{\beta}_{c,1})(z-\widetilde{\beta}_{c,2})}}
  \tanh^{-1}\left(\sqrt{\frac{\widetilde{\beta}-\widetilde{\beta}_{c,2}}{\widetilde{\beta}-\widetilde{\beta}_{c,1}}}\right).
\end{equation}
Therefore we find that the integrand of
\begin{equation} 
\frac{\partial^3 F^{(2)}(\widetilde{T}_c)}{\partial T^3}-\frac{\partial^3 F^{(1)}(\widetilde{T}_c)}{\partial T^3}
\end{equation}
diverges as $(T-\widetilde{T}_c)^{-1}\,(\log(T-\widetilde{T}_c))^{-2}$.
\ack
The financial support of the Universidad Complutense under project GR58/08-910556, the Comisi\'on Interministerial de Ciencia y
Tecnolog\'{\i}a under projects FIS2008-00200 and FIS2008-00209, and the ESF programme MISGAM are gratefully acknowledged.
\section*{Appendix A: The $\tau$-function of the Whitham hierarchy}
In this appendix we prove that the free energy $F$ defined in~(\ref{tau}) satisfies~(\ref{idtau}). The derivative of $F$ with respect to $t_n$ for $n\geq1$ is
\begin{equation}
  \label{a1}
  \partial_{t_n} F = \frac{1}{4\pi\rmi}\oint_{\gamma}z^n \rmd S(z)
                             + \frac{1}{4\pi\rmi}\oint_{\gamma}V(z)\partial_{t_n}(\rmd S(z))+\frac{t_0}{2}\partial_{t_n}v_{1}.
\end{equation}
We analyze separately each of the three terms of this expression. The first one is easily evaluated using~(\ref{sea}):
\begin{equation}
  \label{a2}
  \frac{1}{4\pi\rmi}\oint_{\gamma}z^n \rmd S(z) = \frac{v_{n+1}}{2}.
\end{equation}

The calculation of the remaining terms is more involved. Using~(\ref{es}) we can write the second term in the form
\begin{equation}
  \label{a3}
  \frac{1}{4\pi\rmi}\oint_{\gamma}V(z) \partial_{t_n} (\rmd S(z))
  = \frac{1}{4\pi\rmi}\oint_{\gamma}V(z)\rmd\,\Omega_n
  = \frac{1}{4\pi\rmi}\sum_{m=1}^{2p}t_m \oint_{\gamma}z^m \rmd \Omega_n.
\end{equation}
Now, taking into account~(\ref{ok}) and~(\ref{ake}),
\begin{eqnarray}
  \label{a6}
  \nonumber
  \oint_{\gamma}z^m\,\rmd\,\Omega_n
  &= \oint_{\gamma}(\Omega_m)_{\oplus} \rmd \Omega_n=\oint_{\gamma}(\Omega_m)_{\oplus} \rmd (\Omega_n)_{\ominus}\\
  \nonumber\\
 \nonumber
  &= -\oint_{\gamma}(\Omega_n)_{\ominus} \rmd (\Omega_m)_{\oplus}
  =-\oint_{\gamma}\Omega_n\,\rmd\,(\Omega_m)_{\oplus}\\
  \nonumber\\
  &=-\oint_{\gamma}\Omega_n\,\rmd\,\Omega_m+\oint_{\gamma} \Omega_n\,\rmd\,(\Omega_m)_{\ominus}.
\end{eqnarray}
\begin{figure}
  \includegraphics[width=6cm]{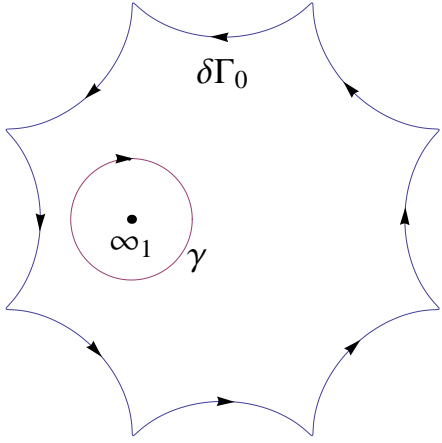}
  \caption{Path $\gamma$.}
\end{figure}
The first integral in the last expression can be transformed into an integral along the boundary $\delta\,\Gamma_0$ of the fundamental domain (see figure~9):
\begin{equation}
\oint_{\gamma}\Omega_n\,\rmd\,\Omega_m=-\oint_{\delta \,\Gamma_0}\Omega_n\,\rmd\,\Omega_m.
\end{equation}
But using the normalization properties and the bilinear relations for Abelian differentials $\rmd \Omega_k$ \cite{spr,far} we have that 
\begin{equation}
\oint_{\delta \,\Gamma_0}\Omega_n\,\rmd\,\Omega_m=\sum_{i=1}^{s-1}(\oint_{a_i} \rmd \Omega_n\,\cdot\,\oint_{b_i} \rmd \Omega_m-\oint_{a_i} \rmd \Omega_m\,\cdot\,\oint_{b_i} \rmd \Omega_n)=0.
\end{equation}
Moreover, it follows easily that
\begin{equation}
  \oint_{\gamma}\Omega_n\,\rmd\,(\Omega_m)_{\ominus}
  =
  \oint_{\gamma}z^n\,\rmd\,(\Omega_m)_{\ominus}=\oint_{\gamma}z^n\,\rmd\,\Omega_m.
\end{equation}
Hence, ({\ref{a6}}) implies that
\begin{equation}
  \label{a7}
  \oint_{\gamma}z^m \rmd \Omega_n= \oint_{\gamma}z^n \rmd \Omega_m,
\end{equation}
and using~(\ref{ds1}) and~(\ref{a3}) we finally obtain for the second term
\begin{eqnarray}
  \label{a8}
  \nonumber
  \frac{1}{4\pi\rmi}\oint_{\gamma} V(z) \partial_{t_n} (\rmd\,S(z))
  &= 
  \frac{1}{4\pi\rmi}\oint_{\gamma}z^n\,\rmd\,S(z)-\frac{t_0}{4\pi\rmi}\,\oint_{\gamma}\Omega_n\,\rmd\,\Omega_0\\
  &=\frac{v_{n+1}}{2} -\frac{t_0}{4\pi\rmi}\,\oint_{\gamma}\Omega_n\,\rmd\,\Omega_0.
\end{eqnarray}

To calculate the third and last term in~(\ref{a1}) note that from~(\ref{uve1}), (\ref{ake}) and~(\ref{o0}) it follows that
\begin{equation}
  \label{a9aa}
  \partial_{t_n} v_1 = L_n = -\frac{1}{2\pi\rmi} \oint_{\gamma'}\Omega_n\,\rmd\,\Omega_0,
\end{equation}
where $\gamma'$ is a large counter clockwise oriented loop which encircles $\bar{J}$ in $\Gamma_2$.  Now~(\ref{a2}), (\ref{a8}) and~(\ref{a9aa}) imply
\begin{equation}
  \partial_{t_n} F = v_{n+1} - \frac{t_0}{4\pi\rmi} \oint_{\gamma+\gamma'}\Omega_n \rmd \Omega_0,
\end{equation}
and integrating along $\delta\Gamma_0$ (see figure~10) and using again~(\ref{nc}), we have
\begin{eqnarray}
  \nonumber
  \oint_{\gamma+\gamma'} \Omega_n \rmd \Omega_0 
  & = -\oint_{\delta\Gamma_0}\Omega_n \rmd \Omega_0\\
  & = -\sum_{i=1}^{s-1}
        \left(\oint_{a_i} \rmd \Omega_n \cdot \oint_{b_i} \rmd \Omega_0-\oint_{a_i} \rmd \Omega_0 \cdot \oint_{b_i} \rmd \Omega_n
        \right) = 0.
\end{eqnarray}
\begin{figure}
  \includegraphics[width=6cm]{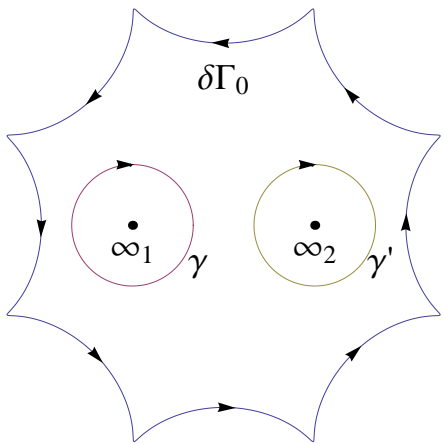}
  \caption{Paths $\gamma$ and $\gamma'$.}
\end{figure}
Therefore we conclude that
\begin{equation}\label{ain}
\partial_{t_n} F=v_{n+1},\quad n\geq 1.
\end{equation}
Consider now the derivative of $F$ with respect to $t_0$. From~(\ref{es}) and~(\ref{uve1}) we have
\begin{equation}
  \label{a9}
  \partial_{t_0} F= \frac{1}{4\pi\rmi}\oint_{\gamma}V(z) \rmd \Omega_0 + \frac{v_1}{2} + \frac{t_0}{2} L_0.
\end{equation}
Moreover, following the same procedure as in the proof of~(\ref{ain}),
\begin{equation}
  \oint_{\gamma} V(z) \rmd \Omega_0 
  = \sum_{n\geq 1}t_n \oint_{\gamma}\Omega_n \rmd \Omega_0
  =-\sum_{n\geq 1}t_n \oint_{\gamma'}\Omega_n \rmd \Omega_0
  = \frac{1}{2} \sum_{n\geq 1}t_n L_n,
\end{equation}
and therefore, using~(\ref{coefv1}) we obtain
\begin{equation}
  \label{aout}
  \partial_{t_0} F= \frac{1}{2} \sum_{n=1}^{2p} t_n L_n + \frac{v_1}{2} + \frac{t_0}{2} L_0=v_1,
\end{equation}
which completes the proof of~(\ref{idtau}).
\section*{Appendix B: Elliptic integrals}
The complete elliptic integrals of first, second and third kind are given by
\begin{eqnarray}
  \label{ellk}
  K(s) = \int_0^1\frac{\rmd t}{\sqrt{1-t^2} \sqrt{1-s\,t^2}},\\
  \label{elle}
  E(s) = \int_0^1\frac{\sqrt{1- s t^2}}{\sqrt{1-t^2}}\rmd t,\\
  \label{ellpi}
  \Pi(r,s) = \int_0^1\frac{\rmd t}{\sqrt{1-t^2} (1-r t^2) \sqrt{1-s\,t^2}},
\end{eqnarray}
respectively. In terms of these integrals,
\begin{equation}
  \fl
  \int_{\beta_2}^{\beta_3} \frac{\rmd x}{\sqrt{(x-\beta_1)(x-\beta_2)(x-\beta_3)(x-\beta_4)}}
  =
  \frac{2 K(s)}{\sqrt{(\beta_3-\beta_1)(\beta_4-\beta_2)}}.
\end{equation}
\begin{equation}
  \fl
  \int_{\beta_2}^{\beta_3} \frac{x \rmd x}{\sqrt{(x-\beta_1)(x-\beta_2)(x-\beta_3)(x-\beta_4)}}
  =
  \frac{2(\beta_4 K(s)+(\beta_3-\beta_4)\Pi(r,s))}{\sqrt{(\beta_3-\beta_1)(\beta_4-\beta_2)}},
\end{equation}
\begin{eqnarray}
  \nonumber
  \fl
  \int_{\beta_2}^{\beta_3} \frac{x^2 \rmd x}{\sqrt{(x-\beta_1)(x-\beta_2)(x-\beta_3)(x-\beta_4)}}\\
  \nonumber
  =
  \frac{1}{\sqrt{(\beta_3-\beta_1)(\beta_4-\beta_2)}}
  \Big(
          (2\beta_4^2-(\beta_4-\beta_3)(\beta_4-\beta_2))K(s)\\
  {} - (\beta_3-\beta_1)(\beta_4-\beta_2) E(s)
     - (\beta_1+\beta_2+\beta_3+\beta_4)(\beta_4-\beta_3)\Pi(r,s)\Big).
\end{eqnarray}

In our study of the merging of two cuts we need the Taylor series of $\Pi(r,s)/K(s)$ near $(r,s)=(0,0)$. From the integral representations~(\ref{ellk}) and~(\ref{ellpi}) it follows immediately that~\cite{wol}
\begin{equation}
  \label{emc}
  K(s) = \frac{\pi}{2}\left(1+\frac{s}{4}+\frac{9 s^2}{64}+\cdots\right),\quad s\rightarrow 0,
\end{equation}
and
\begin{equation}
 \fl
 \label{emc1}
  \Pi(r,s) = \frac{\pi}{2}\left(1+\frac{r}{2}+\frac{s}{4}
                                           +\frac{3 r^2}{8}+\frac{9 s^2}{64}+\frac{3 r s}{16}+\cdots\right),\quad r\rightarrow 0,\,s\rightarrow 0.
\end{equation}
Consequently
\begin{equation}
  \label{pik}
  \frac{\Pi(r,s)}{K(s)} = 1+\frac{1}{2} r+\frac{3}{8} r^2+\frac{1}{16} r s + \cdots,\quad r\rightarrow 0,\,s\rightarrow 0,
\end{equation}
where the remaining  terms are powers  $r^l\,s^k$ of order $l+k\geq 3$.

Similarly, in our study of the birth of a cut we need to estimate $K(s)$ and $\Pi(r,s)$  for $(r,s)$ near $(1,1)$. For $K(s)$ we use the standard formula~\cite{wol}
\begin{equation}
  \label{eek}
  K(s) \sim -\frac{1}{2}\log (1-s),\quad s\rightarrow 1.
\end{equation}
The appropriate asymptotic expression for $\Pi(r,s)$ is obtained from the inequality
\begin{equation}
  \label{eep}
  \fl
  \left|\Pi(r,s) - \frac{\sqrt{2} \theta(r,s)}{\sqrt{1+\sqrt{s}} (1+\sqrt{r})
                                                             \sqrt{1-\sqrt{r}}\,\sqrt{\sqrt{s}-\sqrt{r}}}  \right|
  \leq
  \frac{8\pi}{r^{1/4}\sqrt{\sqrt{s}-\sqrt{r}}},
\end{equation}
where  $0<r<s<1$ and
\begin{equation}
  \theta(r,s) = \tanh^{-1}\left(\sqrt{\frac{\sqrt{s}-\sqrt{r}}{1-\sqrt{r}}}\right).
\end{equation}

To prove~(\ref{eep}) we note that near $t=1$ the integrand in~(\ref{ellpi}) behaves as
\begin{eqnarray}
\nonumber
\fl
 \frac{1}{\sqrt{1-t^2} (1-r t^2) \sqrt{1-s\,t^2}}
 \sim
 & \frac{1}{\sqrt{2} (1+\sqrt{r}) \sqrt{1+\sqrt{s}}} \times\\
 & \frac{1}{\sqrt{1-t} (1-\sqrt{r} t) \sqrt{1-\sqrt{s} t}},
\end{eqnarray}
and that
\begin{equation}
  \int_0^1 \frac{\rmd t}{\sqrt{1-t} (1-\sqrt{r} t) \sqrt{1-\sqrt{s} t}}
  =
  \frac{2 \theta(r,s)}{\sqrt{1-\sqrt{r}} \sqrt{\sqrt{s}-\sqrt{r}}}.
\end{equation}
Furthermore, we have
\begin{eqnarray}
  \label{pii}
  \nonumber\fl
  \Pi(r,s) - \frac{\sqrt{2} \theta(r,s)}{\sqrt{1+\sqrt{s}} (1+\sqrt{r}) \sqrt{1-\sqrt{r}} \sqrt{\sqrt{s}-\sqrt{r}}}
  \\
  = \int_0^1\frac{f(r,s,t)-f(r,s,1)}{\sqrt{1-t} (1-\sqrt{r} t)\,\sqrt{1-\sqrt{s} t}}\rmd t,
\end{eqnarray}
where 
\begin{equation}
 f(r,s,t) = \frac{1}{\sqrt{1+t} (1+\sqrt{r} t) \sqrt{1+\sqrt{s} t}}.
\end{equation}
Now, for all $0\leq r, s, t\leq 1$ we have the bound
\begin{equation}
 \left|\partial_t f(r,s,t)\right| \leq 8
\end{equation}
so that
\begin{equation}
 0 \leq f(r,s,t) - f(r,s,1) \leq 8 (1-t),
\end{equation}
and since
\begin{eqnarray}
  \nonumber
  0 < & \int_0^1 \frac{\sqrt{1-t}}{(1-\sqrt{r} t) \sqrt{1-\sqrt{s} t}} \rmd t\\
  \nonumber
  \leq & \int_0^1 \frac{\rmd t}{(1-\sqrt{r} t) \sqrt{1-\sqrt{s} t}}\\
  \nonumber
  = & \frac{2}{r^{1/4} \sqrt{\sqrt{s}-\sqrt{r}}}
       \left[\tan^{-1}\left(\frac{r^{1/4}}{\sqrt{\sqrt{s}-\sqrt{r}}}\right) -
             \tan^{-1}\left(\frac{r^{1/4}\,\sqrt{1-\sqrt{s}}}{\sqrt{\sqrt{s}-\sqrt{r}}}\right)\right]\\
  \leq & \frac{\pi}{r^{1/4} \sqrt{\sqrt{s}-\sqrt{r}}},
\end{eqnarray}
from~(\ref{pii}) we deduce at once the estimate~(\ref{eep}).
\section*{Appendix C: Asymptotic approximations in the merging and birth of cuts}
In this appendix we  consider the two-cut case only and therefore we dispense with the superscripts.
\subsubsection*{Asymptotic approximations in the merging of two cuts}
We first prove the asymptotic expressions~(\ref{dostres}) for $\beta_2(T)$ and $\beta_3(T)$.  Consider the function
\begin{equation}
  C({\bbeta}) = \beta_4 - (\beta_4-\beta_3) \frac{\Pi(r,s)}{K(s)},
\end{equation}
on a neighborhood 
\begin{equation}
  U = \{{\bbeta}\in\mathbf{R}^4: |\beta_i-\beta_{c,i}|<\delta, i=1,\ldots,4\}
\end{equation}
of the critical point ${\bbeta}_c=(-2,\beta,\beta,2)$ such that $C({\bbeta})$ is analytic in $U$. Given ${\bbeta}\in U$  we use~(\ref{pik}) to approximate $C({\bbeta})$ by its Taylor polynomial at ${\bbeta}_r =(\beta_1,\beta_2,\beta_2,\beta_4)\in U$:
\begin{equation}
  \label{tay}
  \fl
  C({\bbeta}) = \beta_2 
                                 + \frac{1}{2} (\beta_3-\beta_2)
                                 + \frac{1}{8} \frac{1}{\beta_4-\beta_2}\left(1-\frac{1}{2} \frac{\beta_4-\beta_1}{\beta_2-\beta_1}\right)
                                    (\beta_3-\beta_2)^2
                                 +\mathcal{O}((\beta_3-\beta_2)^3).
\end{equation} 
In particular we have 
\begin{equation}
  \left|C({\bbeta})-\beta_2-\frac{1}{2} (\beta_3-\beta_2)\right|
  \leq M (\beta_3-\beta_2)^2,\quad {\bbeta}\in U,
\end{equation}
where
\begin{equation}
  M = \frac{1}{2} \sup_{{\bbeta}\in U}\left|\frac{\partial^2 C}{\partial \beta_3^2}({\bbeta})\right|.
\end{equation} 
Thus, as $T\rightarrow T_c$
\begin{equation}
  \label{cees}
  C({\bbeta})\sim \beta_2+\frac{1}{2}\,(\beta_3-\beta_2).
\end{equation}
Hence, (\ref{merr2}) and~(\ref{cees}) lead to
\begin{eqnarray}
  \label{eb2}
  \dot{\beta}_2 \sim \frac{2}{h(\beta_2,{\bbeta})(\beta_2-\beta_1)(\beta_2-\beta_4)},\\
  \label{eb3}
  \dot{\beta}_3 \sim \frac{2}{h(\beta_3,{\bbeta})(\beta_3-\beta_1)(\beta_3-\beta_4)},
\end{eqnarray}
or, equivalently, to
\begin{eqnarray}
  \label{eb4}
 \left(\frac{1}{2}(\beta_1+\beta_4)+\frac{1}{2}(3\beta_2+\beta_3)-2\beta\right)\dot{\beta}_2 \sim \frac{2}{(\beta^2-4)},\\
 \label{eb5}
 \left(\frac{1}{2}(\beta_1+\beta_4)+\frac{1}{2}(3\beta_3+\beta_2)-2\beta\right)\dot{\beta}_3\sim\frac{2}{(\beta^2-4)}.
\end{eqnarray}
If we now set $T=T_c-t$ in~(\ref{eb4}) and~(\ref{eb5}), substitute  the approximations
\begin{eqnarray}
  \beta_1(T_c-t) \sim -2-\dot{\beta}_1(T_c) t = -2+\frac{t}{(\beta+2)^2},\\
  \beta_4(T_c-t) \sim  2-\dot{\beta}_4(T_c) t = 2-\frac{t}{(\beta-2)^2},
\end{eqnarray}
and look for a solution of the form
\begin{equation}
  \beta_i(T_c-t)\sim \beta-\frac{A_i}{a_i+1}\,t^{a_i+1},\quad -1<a_i<0,\quad i=2,\,3,
\end{equation}
then~(\ref{eb4}) and~(\ref{eb5}) imply
\begin{eqnarray}
  \label{e5}
  \fl
 \left(\frac{1}{(\beta+2)^2} - \frac{1}{(\beta-2)^2}\right) A_2 t^{a_2+1} -
 \left(\frac{3 A_2^2}{a_2+1} t^{2a_2+1} + \frac{A_2\,A_3}{a_3+1} t^{a_2+a_3+1}\right) \sim \frac{4}{\beta^2-4},
 \\
 \fl
 \left(\frac{1}{(\beta+2)^2} - \frac{1}{(\beta-2)^2}\right) A_3 t^{a_3+1} -
 \left(\frac{3 A_3^2}{a_3+1} t^{2a_3+1} + \frac{A_2\,A_3}{a_2+1} t^{a_2+a_3+1}\right) \sim \frac{4}{\beta^2-4}.
\end{eqnarray}
The solution of this system of equations is
\begin{equation}
  \label{e6}
  a_2 = a_3 = -\frac{1}{2},\quad A_2=-A_3= \frac{1}{\sqrt{4-\beta^2}},
\end{equation}
which completes the proof of~(\ref{dostres}).
\subsubsection*{Asymptotic approximations in the birth of a cut}
In our next discussion all the functions involved will be considered as $T$-dependent functions near $\widetilde{T}_c$.  We note that $\beta_4-\beta_3\rightarrow 0$ and
\begin{equation}
  \label{as1}
  \fl
 1-r \sim \frac{\beta_4-\beta_3}{\widetilde{\beta}-\widetilde{\beta}_{c,2}},
 \quad
1-s \sim \frac{\widetilde{\beta}_{c,2}-\widetilde{\beta}_{c,1}}{
                       (\widetilde{\beta}-\widetilde{\beta}_{c,1})(\widetilde{\beta}-\widetilde{\beta}_{c,2})}
                       (\beta_4-\beta_3),
  \quad
  s-r \sim \frac{\beta_4-\beta_3}{\widetilde{\beta}-\widetilde{\beta}_{c,1}}.
\end{equation}
Thus from~(\ref{eek}) we get   
\begin{equation}
  \label{as2}
  K(s) \sim -\frac{1}{2}\,\log(\beta_4-\beta_3).
\end{equation}
Let us look for an asymptotic approximation to $\Pi(r,s)$ near $\widetilde{T}_c$. Taking into account that as $T\rightarrow \widetilde{T}_c$,
\begin{equation}
  (\beta_1,\beta_2,\beta_3,\beta_4)\rightarrow (\widetilde{\beta}_{c,1},\widetilde{\beta}_{c,2},\widetilde{\beta},\widetilde{\beta}),
\end{equation}
\begin{equation}
  r=\frac{\beta_3-\beta_2}{\beta_4-\beta_2}\rightarrow 1,
\end{equation}
\begin{equation}
  s=\frac{\beta_4-\beta_1}{\beta_3-\beta_1} r \rightarrow 1,
\end{equation}
\begin{equation}
  \frac{s-r}{1-r} = \frac{\beta_3-\beta_2}{\beta_3-\beta_1}
                             \rightarrow \frac{\widetilde{\beta}-\widetilde{\beta}_{c,2}}{\widetilde{\beta}-\widetilde{\beta}_{c,1}},
\end{equation}
\begin{equation}
  \frac{\sqrt{s}-\sqrt{r}}{1-\sqrt{r}} = \frac{s-r}{1-r}  \frac{1+\sqrt{r}}{\sqrt{s}+\sqrt{r}}
 \rightarrow \frac{\widetilde{\beta}-\widetilde{\beta}_{c,2}}{\widetilde{\beta}-\widetilde{\beta}_{c,1}},
\end{equation}
\begin{equation}
  \theta(r,s)\rightarrow \tanh^{-1}\left(\sqrt{ \frac{\widetilde{\beta}-\widetilde{\beta}_{c,2}}{\widetilde{\beta}-\widetilde{\beta}_{c,1}}}\right),
\end{equation}
we find that
\begin{eqnarray}
  \nonumber\fl
  \frac{8\pi}{r^{1/4}\sqrt{\sqrt{s}-\sqrt{r}}} \left(\frac{\sqrt{2} \theta(r,s)}{\sqrt{1+\sqrt{s}} (1+\sqrt{r})
                                                             \sqrt{1-\sqrt{r}}\,\sqrt{\sqrt{s}-\sqrt{r}}} \right)^{-1}
 = \\
 \frac{8\pi\sqrt{1+\sqrt{s}}(1+\sqrt{r}) \sqrt{1-\sqrt{r}}}{r^{1/4}\sqrt{2} \theta(r,s)} 
  \rightarrow 0.
\end{eqnarray}
Therefore it follows from~(\ref{eep}) that 
\begin{equation}
  \label{sip}
  \Pi(r,s) \sim \frac{\sqrt{2}\,\theta(r,s)}{
                                \sqrt{1+\sqrt{s}}(1+\sqrt{r})\sqrt{1-\sqrt{r}}\,\sqrt{\sqrt{s}-\sqrt{r}}}.
\end{equation}
Moreover, we have that
\begin{equation}
  \sqrt{1-\sqrt{r}} \sim \sqrt{\frac{1-r}{2}},
  \quad
  \sqrt{\sqrt{s}-\sqrt{r}} \sim \sqrt{\frac{s-r}{2}},
\end{equation}
so that~(\ref{sip}) can be rewritten as
\begin{equation}
  \label{sip2}
  \Pi(r,s)\sim\frac{\tanh^{-1}(\sqrt{\frac{s-r}{1-r}})}{\sqrt{s-r}\sqrt{1-r}}.
\end{equation}
Then, using~(\ref{as1}) it follows that
\begin{equation}
  \label{as3}
  \Pi(r,s) \sim \frac{\sqrt{(\widetilde{\beta}-\widetilde{\beta}_{c,1})(\widetilde{\beta}-\widetilde{\beta}_{c,2})}}{\beta_4-\beta_3}
                       \tanh^{-1}\left(\sqrt{\frac{\widetilde{\beta}-\widetilde{\beta}_{c,2}}{\widetilde{\beta}-\widetilde{\beta}_{c,1}}}\right),\end{equation}
which together with~(\ref{as2}) lead to
\begin{equation}
  \label{bc111}
  \frac{\Pi(r,s)}{K(s)}
  \sim
  -\frac{2\,\sqrt{(\widetilde{\beta}-\widetilde{\beta}_{c,1})(\widetilde{\beta}-\widetilde{\beta}_{c,2})}
           }{(\beta_4-\beta_3)\,\log(\beta_4-\beta_3)}
  \tanh^{-1}\left(\sqrt{\frac{\widetilde{\beta}-\widetilde{\beta}_{c,2}}{\widetilde{\beta}-\widetilde{\beta}_{c,1}}}\right).
\end{equation}

\end{document}